# Multi-technology co-optimization approach for sustainable hydrogen and electricity supply chains considering variability and demand scale

Sunwoo Kim[1,2], Joungho Park[3], and Jay H. Lee[1*]

[1] Mork Family Department of Chemical Engineering and Material Sciences, University of Southern California, 925 Bloom Walk, Los Angeles, CA 90089, USA

[2] Department of Chemical and Biomolecular Engineering, Korea Advanced Institute of Science and Technology (KAIST), 291, Daehak-ro, Yuseong-gu, Daejeon, 34141, Republic of Korea

[3] Energy AI & Computational Science Laboratory, Korea Institute of Energy Research, 71-2 Jang-dong, Yuseong-gu, Daejeon, 305-343, Republic of Korea

[*] Corresponding author's E-mail: jlee4140@usc.edu

---

In the pursuit of a carbon-neutral future, hydrogen emerges as a pivotal element, serving as a carbon-free energy carrier and feedstock. As efforts to decarbonize sectors such as heating and transportation intensify, understanding and navigating through the dynamics of hydrogen demand expansion becomes critical. Transitioning to hydrogen economy is complicated by varying regional scales and types of hydrogen demand, with forecasts indicating a rise in variable demand that calls for diverse production technologies. Currently, steam methane reforming is prevalent, but its significant carbon emissions make a shift to cleaner alternatives like blue and green hydrogen imperative. Each production method possesses distinct characteristics, necessitating a thorough exploration and co-optimization with electricity supply chains as well as carbon capture, utilization, and storage systems. Our study fills existing research gaps by introducing a superstructure optimization framework that accommodates various demand scenarios and technologies. Through case studies in California, we underscore the critical role of demand profiles in shaping the optimal configurations and economics of supply chains and emphasize the need for diversified portfolios and co-optimization to facilitate sustainable energy transitions.

---

## Graphical abstracts

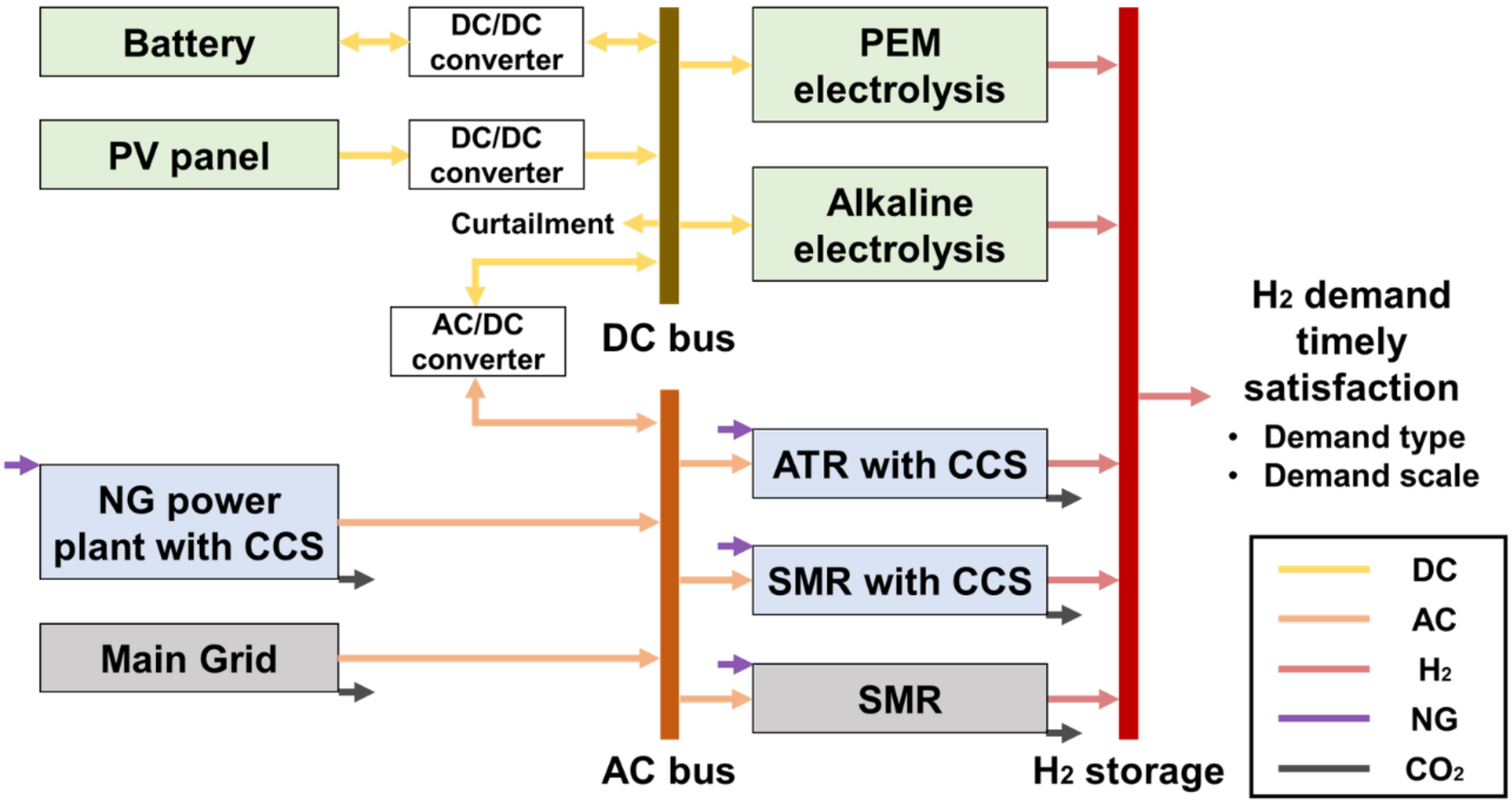

- Navigating the transition of **multi-technology hydrogen** and **electricity supply chain**
- Analyzing **optimal configuration** and **economic feasibility** by varying **demand type & scale**

## Main

In the global pursuit of a carbon-neutral future, hydrogen is recognized as a pivotal element [1-5], with projections indicating a substantial rise in demand, particularly for heating and transportation. This trend underscores the need for a nuanced understanding of hydrogen dynamics [6, 7] as the varied regional demands highlight the complexity of transitioning toward sustainable energy systems [4, 8]. Despite current steady demand, forecasts indicate an increasing variability in future hydrogen requirements [9-11], necessitating a thorough exploration of diverse hydrogen production technologies, each with unique characteristics and implications. However, this urgency is often overshadowed by previous research gaps.

Multiple technologies for hydrogen production are available, with steam-methane reforming (SMR) being particularly notable for its efficiency in using electricity and natural gas (NG) [8, 12]. However, the high carbon dioxide emissions from SMR demand a shift to greener alternatives [13]. Among these, blue hydrogen involves capturing a substantial portion of $CO_2$ emissions during NG-based hydrogen production, albeit at the cost of increased electricity and NG consumption [14, 15]. In contrast, green hydrogen, produced through water electrolysis using electricity, is only environmentally beneficial when the electricity comes from carbon-free sources[13, 16, 17]. Each hydrogen production method offers different economies of scale, operational flexibility, energy consumption rates, and levels of carbon emissions. Furthermore, the interplay between hydrogen and electricity supply chains underscores the need for co-optimization to efficiently meet evolving demand dynamics [18, 19]. Options in the electricity supply chain include solar power, which, while emission-free, faces production regulation challenges, and low carbon (low C) electricity, which, though flexible, comes with high installation costs and residual emissions [20, 21].

Previous studies have often missed critical aspects such as the benefits of diversified portfolios for hydrogen production and the need for co-optimized supply chain configurations. Our research addresses these gaps by developing a superstructure optimization framework that simultaneously evaluates various hydrogen demand scenarios and production technologies within a unified optimization model. This methodology is crucial for identifying optimal strategies amid expanding demand and diverse technological environments, emphasizing the advantages of diversified portfolios for both hydrogen and electricity supply chains. Such diversification enhances the energy system's resilience and adaptability to fluctuating demands and renewable sources, playing a vital role in a sustainable energy transition. Moreover, our study underscores the importance of co-optimizing hydrogen and electricity supply chains to efficiently meet hydrogen demands, an aspect often neglected in previous research. Understanding the interplay and mutual influence of these supply chains is essential for creating effective sustainable energy production and distribution strategies. Our analysis also highlights the crucial role of resilient electricity supply chains, particularly in supporting the energy requirements of greener hydrogen production methods, making them fundamental to a sustainable energy framework.

Through case studies in California, a leader in sustainable energy initiatives [22-24], we explore how varying demand scales and types affect the economics and optimal configuration of hydrogen and electricity supply chains. By exploring the co-optimization of these chains and the potential of diversified portfolios, we reveal complex interactions that provide comprehensive insights for policymakers and industry stakeholders. Ultimately, our findings contribute to a deeper understanding of the complexities involved in moving towards a greener, more resilient energy ecosystem, with implications that reach beyond California to influence global sustainable energy discussions.

## Superstructure for co-optimizing hydrogen and electricity supply chain

As illustrated in Fig. 1, the carbon intensity of hydrogen production varies significantly based on the electricity source, even when employing the same hydrogen production technology. For SMR, this variation is minimal due to its low electricity consumption rate of approximately 1 MWh per ton of hydrogen. However, the variation is more pronounced for blue hydrogen, which involves CCS alongside SMR and ATR, resulting in an electricity consumption rate about four times higher. The variation is exceptionally significant for green hydrogen, with a rate approximately fifty times higher.

Despite its higher energy consumption, green hydrogen has the greatest potential to reduce carbon intensity when the electricity used is sourced from renewable energy. Therefore, it is crucial to optimize both the hydrogen and electricity supply chains simultaneously to address environmental considerations. This optimization not only improves environmental outcomes but also has a substantially impact on the economic viability of hydrogen production, as detailed in the results and discussion section.

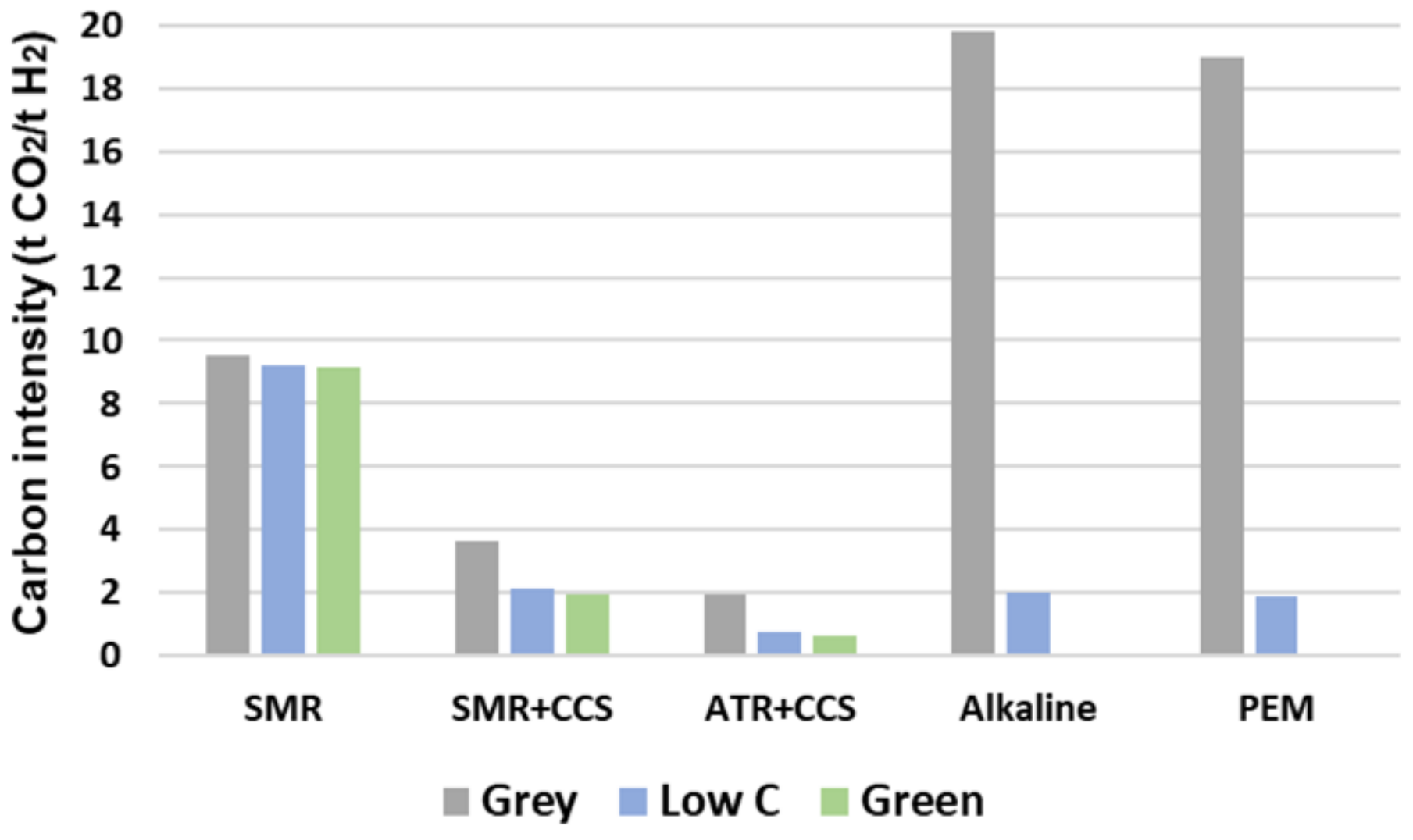

Fig. 1. Carbon intensity for various hydrogen production technologies using different electricity power sources: steam methane reformers (SMR), carbon capture and storage (CCS) technology, autothermal reformers (ATR), alkaline electrolyzers, and proton exchange membrane (PEM) electrolyzers.

In Fig. 2, the electricity supply sources include a photovoltaic (PV) panel with a battery, an NG power plant equipped with carbon capture and storage (CCS), and the main grid. The NG power plant, featuring CCS, is highlighted as a low C option for electricity generation. It benefits from requiring minimal space and leveraging well-established infrastructure for NG in many regions, which simplifies transportation, storage, and utilization. Given the varied nature of each electricity source, appropriate converters and buses are essential to integrate these power sources effectively. Further details are provided in the Supplementary Materials.

This study also evaluates five hydrogen production technologies: Proton Exchange Membrane (PEM) electrolysis, Alkaline electrolysis, Autothermal Reformer (ATR) with CCS, SMR with CCS, and SMR alone. Each technology operates above a specific minimum threshold and exhibits distinct techno-economic characteristics, including rates of electricity and NG consumption, $CO_2$ emissions, and scale economies. Additional information on these technologies is provided in the Supplementary Materials.

This study focuses on meeting regional hydrogen demand promptly, considering both the types and scales of demand. Hydrogen demand is classified into two types: constant and variable. While current hydrogen demand is predominantly constant, future demand is expected to be more variable. The demand scale varies from 2 kt/year to 200 kt/year, each with its own economic feasibility and optimal configurations that depend on both the scale and type of demand [25]. To address this, our analysis covers two types of hydrogen demand (constant and variable) and three demand scales (2, 20, and 200 kt/year) to identify the most effective hydrogen and electricity supply networks and assess their economic viability. Detailed information is available in the Supplementary Material.

Additionally, considering the annual advancements in renewable energy technologies, batteries, and water electrolyzers, which have led to price reductions, along with the impact of increasing carbon taxes, we observed changes in the optimal configurations of hydrogen and electricity supply networks. These changes are detailed in Table 1.

To provide a comprehensive understanding of the energy transition, we also explored various scenarios, summarized in Table 2. Initially, we analyze the benefits of a diversified portfolio by comparing the results of the Unique case, which employs a single hydrogen production technology and electricity source, to those of the Base case, which involves co-optimizing a diversified hydrogen and electricity supply chain. We then examine how the advantages of a diversified portfolio change in the absence of low C electricity, which can occur due to geographic constraints that may hinder the installation of NG power plants with CCS. Furthermore, we analyze how the benefits of a diversified portfolio vary under conditions where NG prices are tripled. Given the high volatility of NG prices in California, we assess the impact of this price variability. Lastly, we examine the effect of a carbon tax on the optimal supply chain to provide insights for policymakers.

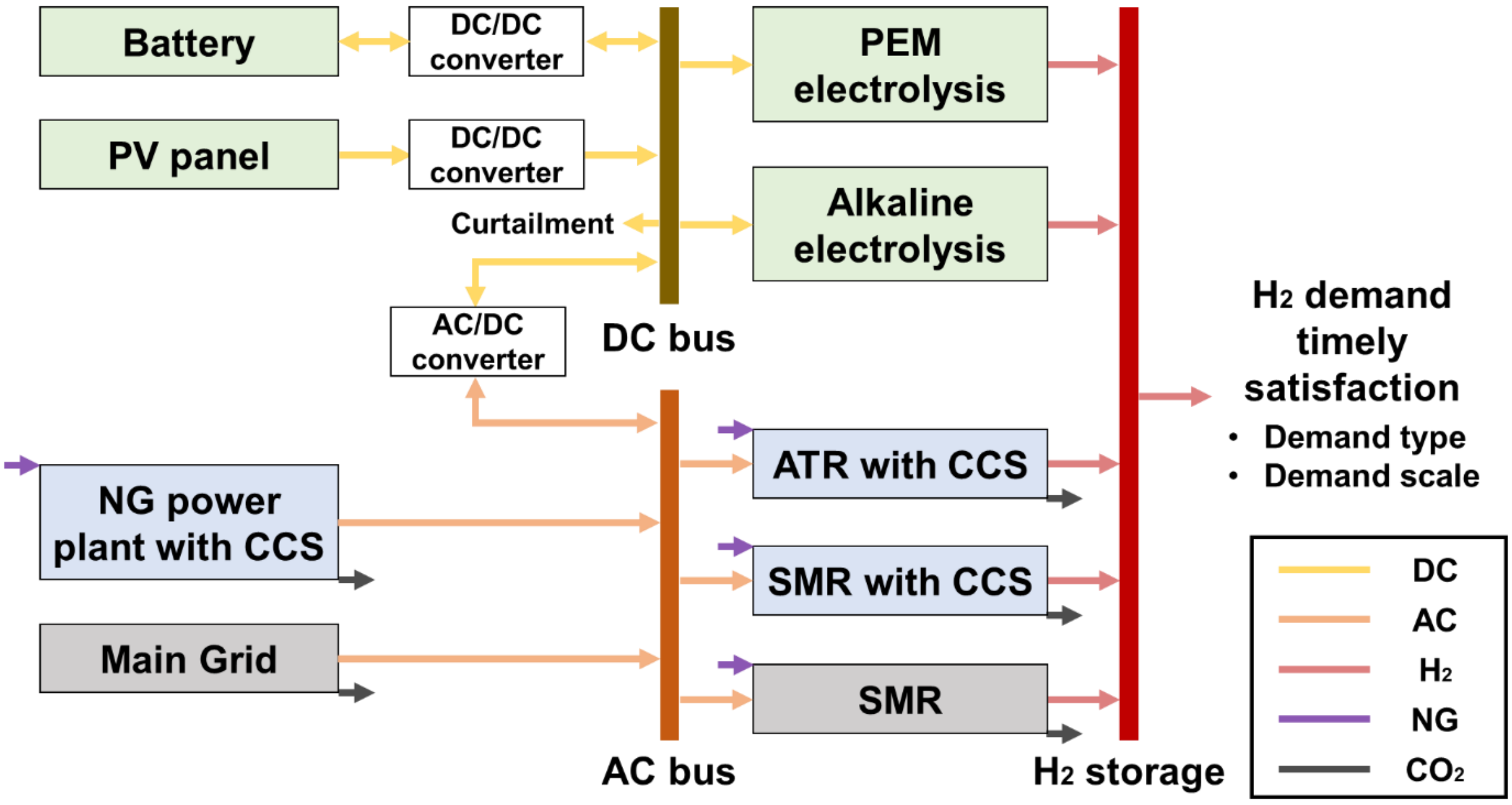

Fig. 2. Description of the Integrated Supply Chain System: This system integrates multiple technologies to ensure the timely supply of hydrogen ($H_2$) across different types and scales. Key components include photovoltaic (PV) systems, natural gas (NG) sources, carbon capture and storage (CCS) technology, proton exchange membrane (PEM) electrolyzers, autothermal reformers (ATR), and steam methane reformers (SMR). The system also incorporates direct current (DC) and alternating current (AC) electrical systems to efficiently manage and distribute energy.

Table 1. Summary of investment costs for renewable energy facilities and $CO_2$ tax trajectory, 2030-2050.

| | **2030** | **2040** | **2050** |
|---|---|---|---|
| **Solar PV panel** | $751/kW | $685/kW | $618/kW |
| **Battery** | $224/kWh | $196/kWh | $168/kWh |
| **Alkaline electrolyzer** | $550/kW | $413/kW | $330/kW |
| **PEM electrolyzer** | $789/kW | $562/kW | $437/kW |
| **Averaged $CO_2$ tax** | $100/t $CO_2$ | $150/t $CO_2$ | $200/t $CO_2$ |

Table 2. Case study overview: This table provides a summary of each case study's name along with a detailed description corresponding to each case.

| Case name | Description |
|---|---|
| **Unique case** | Unique hydrogen technology is adopted while utilizing a single electricity power source. |
| **Base case** | Standard scenario for comparison. |
| **No low C electricity case** | Scenario where a natural gas (NG) power plant with carbon capture and storage (CCS) is absent. |
| **Expensive NG case** | Natural gas (NG) prices are assumed to triple compared to the base case, rising from \$3.5/MMBtu to \$10.5/MMBtu. |
| **$CO_2$ tax sensitivity analysis case** | Analyzes the impact of varying the $CO_2$ tax from -50 to +50% relative to the base case value in 2050. |

**Evaluation metrics**

Our model is designed to minimize the Levelized Cost of Hydrogen (LCOH) while ensuring the timely fulfillment of hydrogen demand. The key economic metrics are represented by LCOH and the Levelized Cost of Electricity (LCOE):

$$LCOH = \frac{\text{Total annualized cost}}{\text{Annual H}_2\text{ demand}} = LCOH^{\text{CO}_2\text{tax}} + LCOH^{\text{Electricity}} + LCOH^{\text{Natural gas}} + LCOH^{\text{Facility}} \quad \text{(Equation 1)}$$

$$LCOE = \frac{\text{Total annualized cost for electricity}}{\text{Annual net electricity demand}} \quad \text{(Equation 2)}$$

These metrics evaluate the economics of the overall system and the electricity supply chain specifically.

- LCOH encompasses various cost components including $CO_2$ tax ( $LCOH^{\text{CO}_2\text{tax}}$ ), electricity ( $LCOH^{\text{Electricity}}$ ), natural gas ( $LCOH^{\text{Natural gas}}$ ), and facility costs ( $LCOH^{\text{Facility}}$ ).
- LCOE is calculated by dividing the total annualized cost of electricity production, required to meet hydrogen demand, by the amount of electricity produced.

The detailed equations and techno-economic parameters used in these calculations are provided in the Supplementary Materials.

**Unique case: Unique hydrogen technology using a single electricity power source.**

In this scenario, we evaluate the economic viability by adapting a unique hydrogen technology with a single electricity power source. We explore the optimal hydrogen and electricity supply chain by taking into account the hourly and seasonal variability of renewable energy sources. The analysis distinguishes between constant and variable hydrogen demand scenarios:

- Constant demand: Assumes a steady hydrogen demand throughout.
- Variable demand: Hydrogen demand fluctuates in accordance with the hourly seasonal variability of electricity demand.

Results are presented in Fig. 3. Key findings include:

- The variable hydrogen demand scenario results in an LCOH that is \$0.3-\$0.7/kg $H_2$ more expensive than the constant demand scenario. This increase is primarily attributed to higher facility costs, necessitated by larger installation sizes of hydrogen tank and hydrogen producing facility to manage the volatile hydrogen demand.
- Reforming technologies rely on low C electricity because it supports a consistently high utilization rate, which is optimal given the high installation costs. Low C electricity can produce power uniformly, unlike renewable energy, which has high variability
- For constant demand, the LCOE of low C electricity remains stable regardless of demand scale. However, for variable demand, the LCOE of low C electricity increases with higher demand. This is because the high installation cost of the reformer is optimized by operating small plants at high utilization rates and storing hydrogen in tanks. Small-sized NG power plants with CCS are also used to maintain high utilization. Conversely, as demand scales up, the LCOE increases due to economies of scale making the reformer cheaper, leading to the installation of larger plants and larger NG power plants with CCS.
- Notably, for a 2 kt hydrogen demand in 2050, a SMR is used for constant demand, while a PEM electrolyzer is chosen for variable demand. Water electrolyzers have lower installation costs compared to blue hydrogen and SMR facilities, which require uniformly high utilization rates. Consequently, reforming technology is preferable for constant demand, whereas green hydrogen is more favorable under variable demand conditions.
- The production cost of electricity is lower for green electricity compared to low C electricity. However, the LCOE increases as much as the produced electricity is curtailed due to intermittency. Consequently, for reforming technologies that require a high operating rate, the LCOE of renewable energy is very high. In contrast, for green hydrogen technology, the LCOE of renewable energy is lower due to reduced installation costs and greater operational flexibility.

We conducted a detailed analysis of the relationship between hydrogen production technologies and various electricity options as illustrated in Fig. 4.

For green hydrogen technology,

- In 2030, when renewable energy technologies are still developing, the most economical approach is to use low C electricity to produce hydrogen via an alkaline water electrolyzer. This method yields a LCOH of approximately \$4.4/kg $H_2$ and a LCOE of about \$78/MWh. By 2050, the optimal economic scenario involves using PEM electrolyzers powered by green electricity, achieving an LCOH of about \$4.3/kg $H_2$ and an LCOE of \$54/MWh. The PEM electrolyzer's advantage lies in its ability to operate reliably with fewer batteries compared to the alkaline electrolyzer, due to its lower minimum load requirement.
- For variable demand, the preference for green electricity over low C electricity increases. This is because the economics of using green electricity remain stable with changing demand types, whereas the cost of utilizing low C electricity rises more significantly under variable demand conditions.

For reforming technology,

- Low C electricity proves to be more economical across all scenarios. The LCOE with green electricity is consistently more than \$100/MWh higher than that with low C electricity.
- Notably, for constant demand, the LCOE of green electricity tends to decrease as demand scale increases, due to economies of scale reducing the installation cost of the reforming plant.
- Conversely, for variable demand, the LCOE increases with demand scale. This increase is driven by a trade-off between the size of the hydrogen tank and the rest of the facilities. As demand scale grows, the optimization strategy shifts towards smaller hydrogen tanks and larger production capacities to handle demand volatility. Thus, there is a competition between expanding hydrogen tanks and enhancing production capacity to manage variable demand, with a tendency to favor increased production capacity as demand scale rises.

The detailed design specifications are available in the Supplementary Materials.

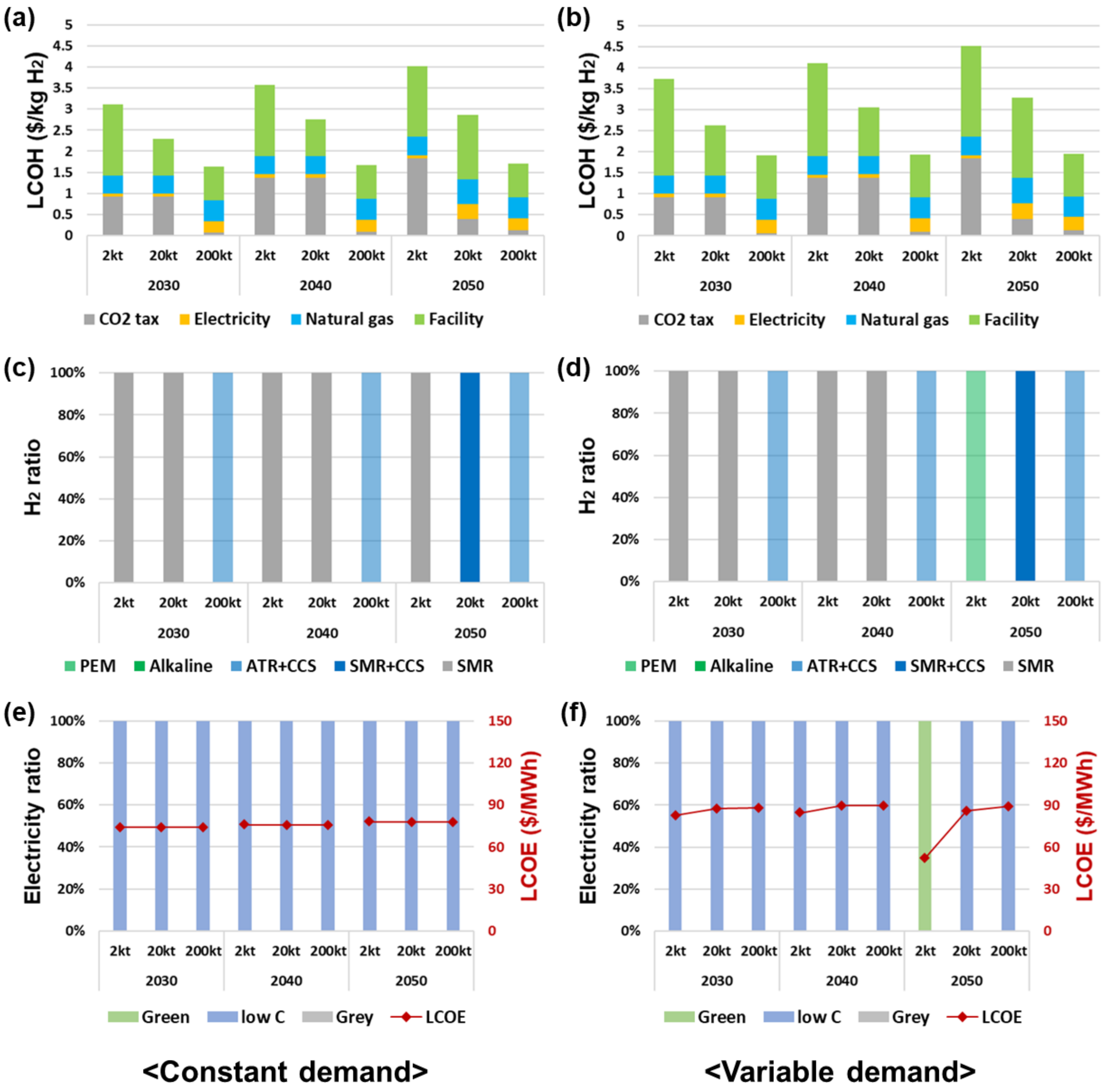

Fig. 3. Unique case: (a) LCOH breakdown for constant demand: This chart details the LCOH components including $CO_2$ tax, electricity, NG, and facility costs for constant hydrogen demand, varying across different demand scales from 2030 to 2050. (b) LCOH breakdown for variable demand: Similar to (a), this chart displays the LCOH composition for variable hydrogen demand scenarios. (c) Hydrogen production technology ratio for constant demand: This graph shows the distribution of hydrogen production technologies—PEM electrolysis, alkaline electrolysis, ATR+CCS, SMR+CCS, and SMR—for a constant demand scale, observed from 2030 to 2050. (d) Hydrogen production technology ratio for variable demand: Similar to (c), this graph compares the technology distribution for variable hydrogen demand. (e) Electricity ratio and LCOE for constant demand: This section analyzes the proportion of electricity consumption and the corresponding LCOE for different hydrogen production scales under constant demand conditions from 2030 to 2050. (f) Electricity ratio and LCOE for variable demand: Similarly, this part examines how the variability in hydrogen demand affects the electricity consumption ratios and LCOE.

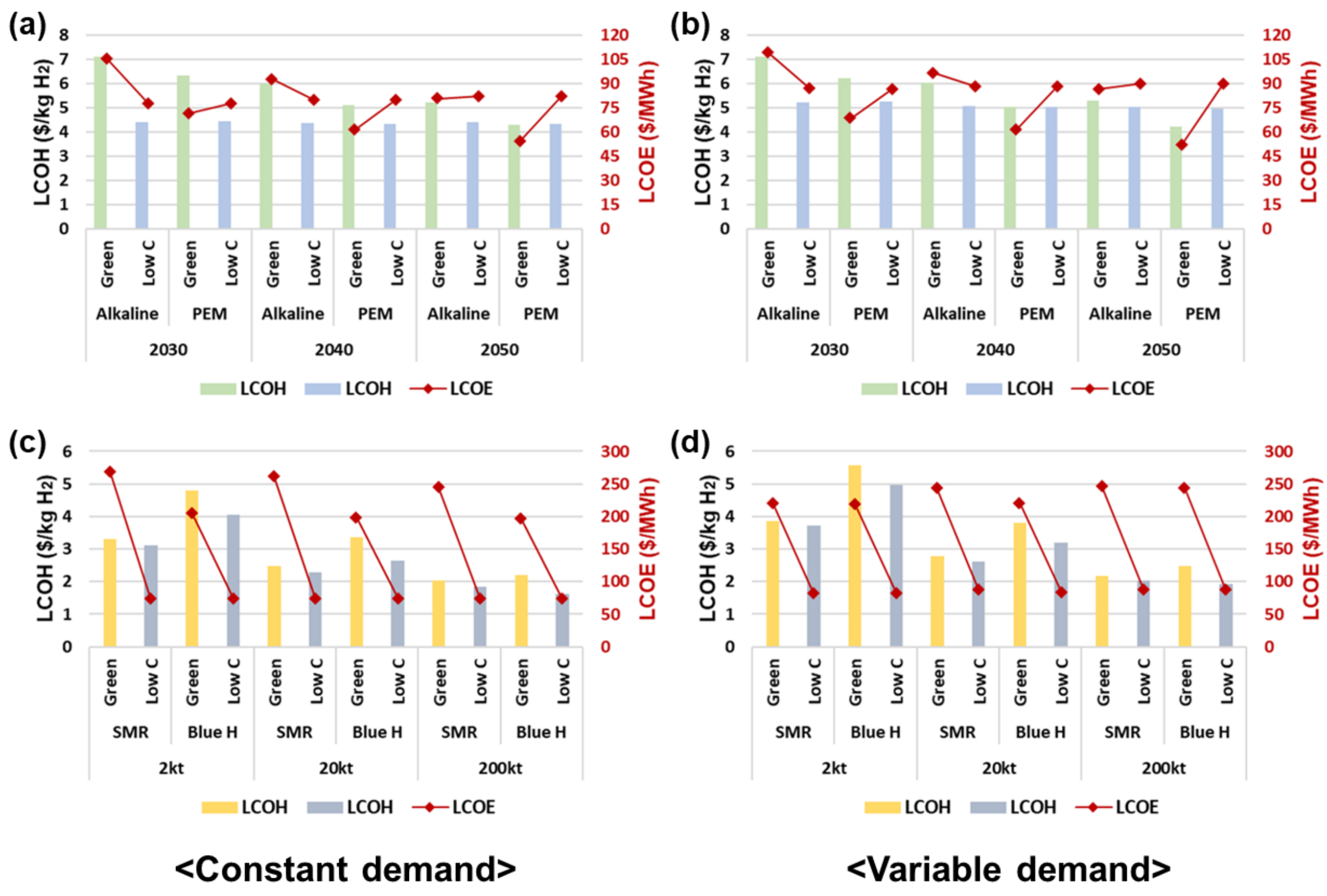

Fig. 4. LCOH and LCOE for Green hydrogen technologies across different electricity sources and years: (a) Constant demand, (b) Variable demand, LCOH and LCOE for Reforming technologies across different electricity sources and demand scale in 2030: (c) Constant demand, (d) Variable demand.

## Base case: Diversified hydrogen and electricity portfolio

In this section, we explore the optimal diversified portfolio of the hydrogen and electricity supply chain, considering the hourly and seasonal variability of renewable energy sources. Fig. 5 presents the LCOH and carbon intensity for both the Unique case and the Base case while Fig. 6 illustrated the optimal result for the base case.

Key findings in LCOH perspective in Fig. 5:

- For constant demand, there is a minor reduction in the LCOH, typically less than 1%, except for the scenario involving a demand of 2 kt in 2050. In this specific case, the LCOH decreases significantly by approximately 7%, equivalent to around \$0.3/kg $H_2$.

- For variable demand, the LCOH reduction is more pronounced, around 4%. The reduction is particularly substantial at the 2 kt hydrogen demand level in 2050, where the LCOH decreases by about 18%, translating to a cost reduction of \$0.8/kg $H_2$.
- The diversified portfolio shows greater benefits in variable demand cases, may due to the high operating flexibility of green hydrogen technologies.

Key findings in carbon intensity perspective in Fig. 5:

- For constant demand, the reduction in carbon intensity is generally modest, less than 3%, except for the 2 kt scenario in 2050. In this instance, the carbon intensity drops dramatically by 64%, equivalent to a reduction of approximately 5.9 t $CO_2$/t $H_2$.
- For variable demand, the average reduction in carbon intensity is about 7%, with the 2 kt case in 2050 presenting a unique situation. Here, the Unique case shows zero carbon emissions, whereas the Base case indicates an increase in carbon intensity of approximately 1.8 t $CO_2$/t $H_2$.
- While a diversified portfolio does not inherently ensure a reduction in carbon intensity, it generally leads to significant carbon reductions in most scenarios.

Results are presented in Fig. 6. Key findings include:

- The variable hydrogen demand scenario results in an LCOH that is \$0.3-\$0.5/kg $H_2$ more expensive than the constant demand scenario, except for the 2 kt in 2050. This increase is primarily attributed to higher facility costs, necessitated by larger installation sizes to manage the volatile hydrogen demand.
- Notably, for a 2 kt hydrogen demand in 2050, variable demand case shows marginally better economics than constant demand case. This advantage is due to the alignment of hydrogen demand peaks with solar energy production patterns—higher during the day and tapering off at night. This synchronicity benefits green hydrogen operations, which leverage operational flexibility and lower production costs of green electricity.
- Higher electricity prices notably affect the hydrogen supply chain dynamics. By 2040, for small and medium-scale hydrogen demands, SMR, which utilizes minimal electricity, remains favored. However, for large-scale hydrogen production, ATR+CCS based blue hydrogen remains the preferred option due to its significant economic advantages.

In terms of technology preference,

- For variable demand, the shift towards green hydrogen is driven by its lower installed costs and greater operational flexibility relative to blue and grey hydrogen technologies, which perform better under steady, high utilization conditions.
- PEM electrolysis is favored over alkaline electrolysis due to its lower minimum operating constraint (5% compared to 20%), allowing for more flexible operation despite higher installation and operational costs. This flexibility becomes a critical factor when dealing with the variable electricity supply, overturning the previous preference for alkaline electrolysis under constant electricity assumptions.

Technology mix for different demand scenarios can be summarized as follows:

- For constant demand in 2050, the optimal supply chain for small-scale hydrogen includes a mix of PEM, Alkaline, and SMR technologies.
- For variable demand, there is an increased preference for a mix of green hydrogen along with blue or grey hydrogen technologies. This diversification offers advantages as the volatility in electricity supply and hydrogen demand increases, leading to a lower LCOE through increased reliance on green electricity.

In 2050, for medium-scale hydrogen demand, SMR with CCS emerges as the dominant technology over ATR with CCS, which, despite its high energy efficiency, has higher installation and operating costs not justified by its scale benefits. Interestingly, blue and grey hydrogen serve similar functions but do not integrate well in the same energy mix, with one technology typically dominating. Conversely, a blend of PEM electrolysis and alkaline electrolysis emerges as optimal for green hydrogen in certain scenarios, reflecting their complementary strengths in variable supply conditions.

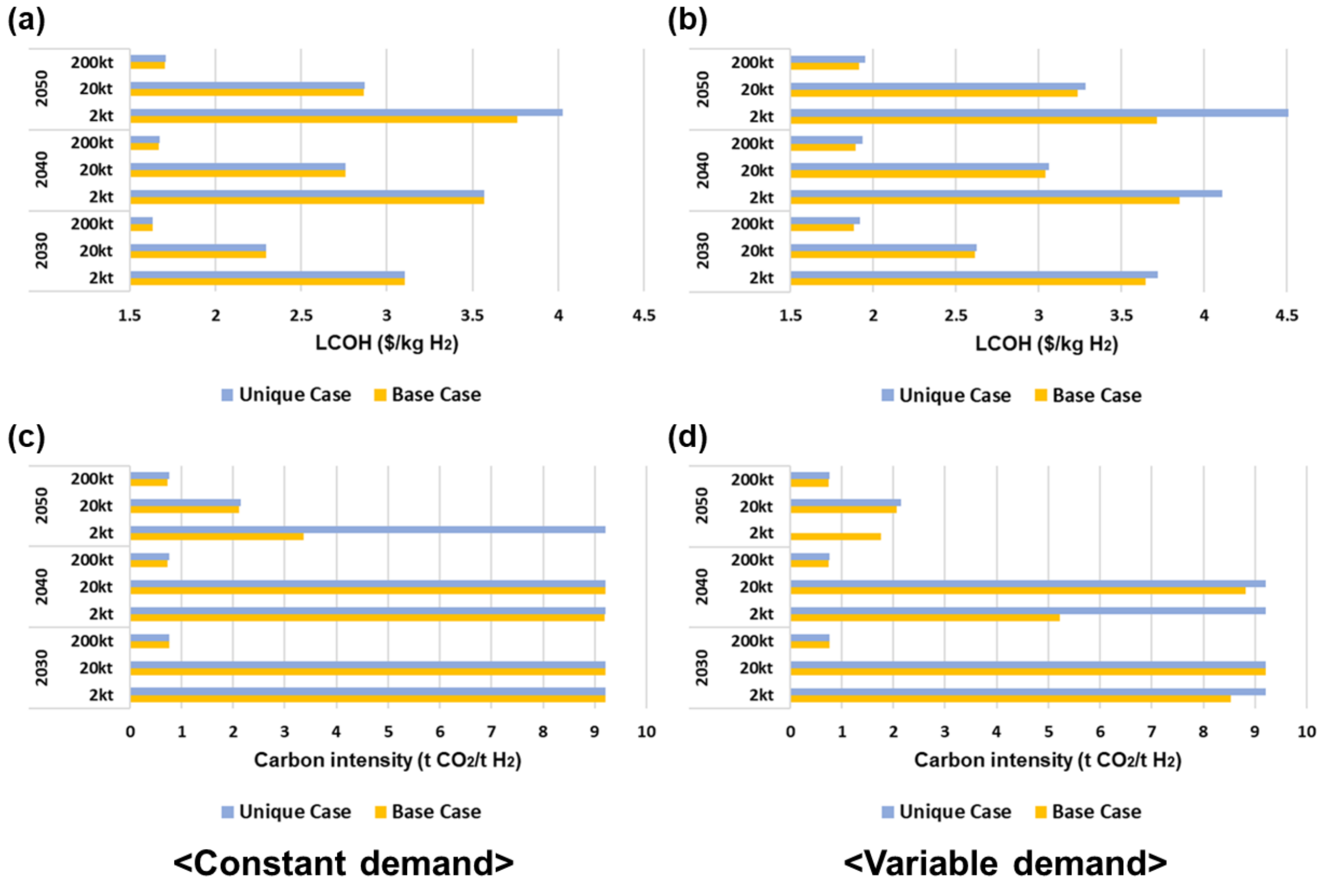

Fig. 5. LCOH for Unique and Base cases: (a) Constant demand, (b) Variable demand, Carbon intensity for Unique and Base cases: (c) Constant demand. (d) Variable demand (Blue bar represents the Unique case, and Orange bar denotes the Base case).

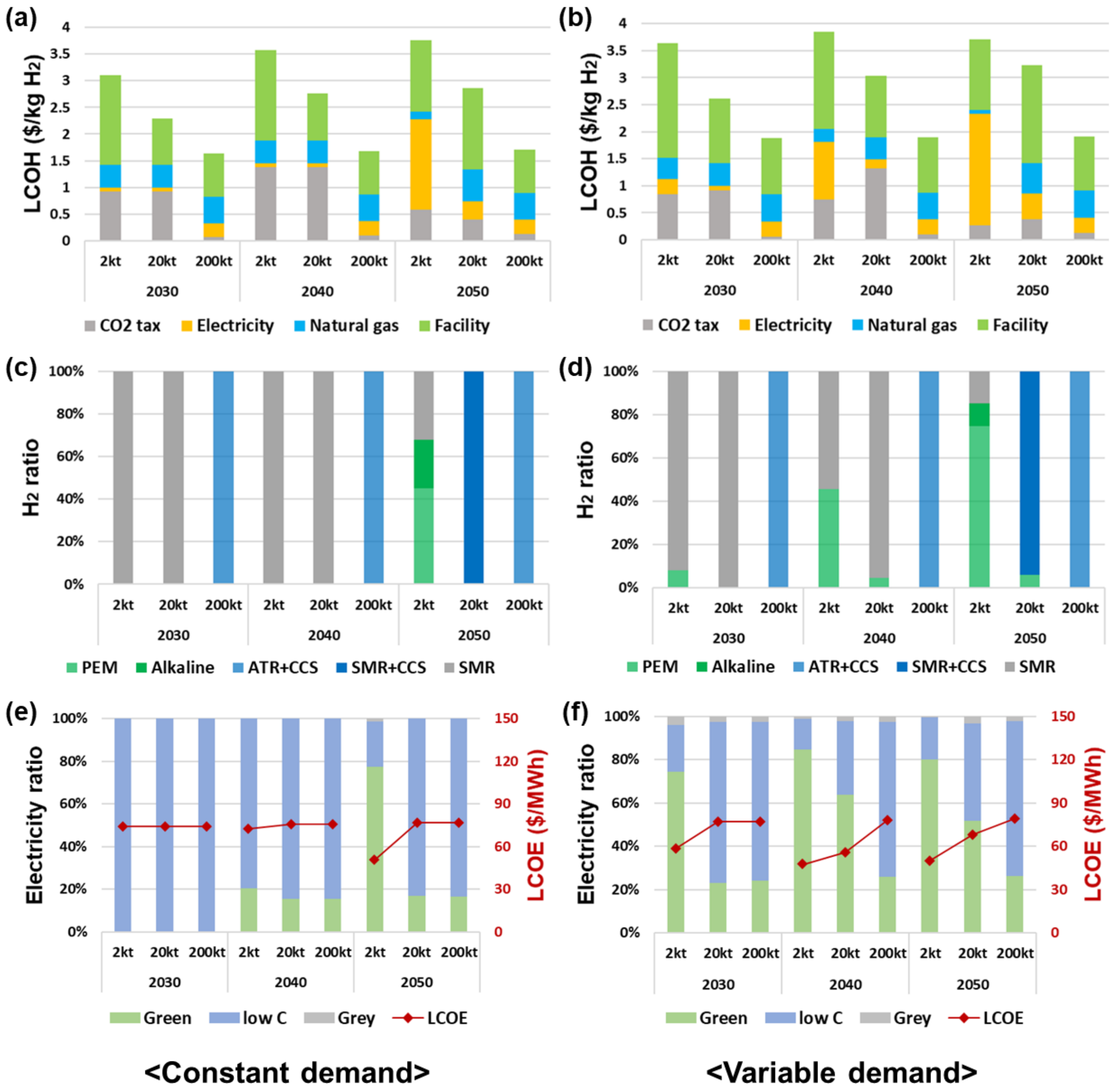

Fig. 6. Base case: (a) LCOH breakdown for constant demand. (b) LCOH breakdown for variable demand. (c) Hydrogen production technology ratio for constant demand. (d) Hydrogen production technology ratio for variable demand. (e) Electricity ratio and LCOE for constant demand. (f) Electricity ratio and LCOE for variable demand.

**No low C electricity option case: Influence of absence of low C electricity option**

In this analysis, we explored the effects on the hydrogen and electricity supply chain when low C electricity—an option for generating electricity with minimal carbon emissions on demand—is unavailable. Fig. 5 presents the LCOH and carbon intensity for both the Unique case and the Base case while Fig. 6 illustrated the optimal result for the base case.

Key findings in LCOH perspective in Fig. 7:

- For constant demand, the LCOH for a diversified portfolio decreased by an average of 6% compared to a unique portfolio. The improvements ranged from \$0.1/kg $H_2$ to \$0.26/kg $H_2$, corresponding to reductions of 3.5% to 11%.
- For variable demand, the LCOH reduction is more pronounced, around 7%. The improvements ranged from \$0.1/kg $H_2$ to \$0.35/kg $H_2$, corresponding to reductions of 4% to 13%.
- This scenario highlights the increasing importance of diversifying electricity production, as the removal of low-cost, controlled production options underscores the benefits of low C electricity.

Key findings in carbon intensity perspective in Fig. 7:

- For constant demand, the majority of carbon intensity reductions were observed, with an average decrease of 1.5 t $CO_2$/t $H_2$, equivalent to about 13%. The most significant reductions were seen in the 200 kt scenario in 2030 and the 2 kt scenario in 2050, with decreases of 5 t $CO_2$/t $H_2$ (equates to ~55%) and 8.4 t $CO_2$/t $H_2$ (equates to ~92%), respectively.
- Variable demand also generally resulted in reduced carbon intensity, with an average decrease of 0.9 t $CO_2$/t $H_2$. However, a notable exception was observed in the 2 kt scenario in 2050, where carbon intensity increased from 0 to 2.8 t $CO_2$/t $H_2$.

Results are illustrated in Fig. 8 and notable observations are summarized below:

- Compared to the Base case, the LCOH increases by approximately \$0.2/kg $H_2$.
- The unit cost of electricity experiences a significant rise, moving from \$50-80/MWh in the Base case to \$50-150/MWh. This substantial increase in electricity costs has led to a greater reliance on SMR-based hydrogen within the supply chain due to its lower electricity consumption.
- Concurrently, there is a noticeable decline in the preference for green hydrogen production. This shift is largely due to the increased electricity costs, which undermine the economic viability of technologies like alkaline electrolysis that have higher operational inflexibilities (minimum operating constraint of 20%).
- The absence of low C electricity options forces a heavier dependence on more expensive grey electricity, thereby raising overall electricity production costs and posing significant challenges to the adoption and expansion of green hydrogen solutions.

This scenario highlights the critical role that low carbon electricity plays in enabling competitive green hydrogen production. Without it, the supply chain leans towards technologies that are less dependent on electricity, slowing progress towards more sustainable energy solutions.

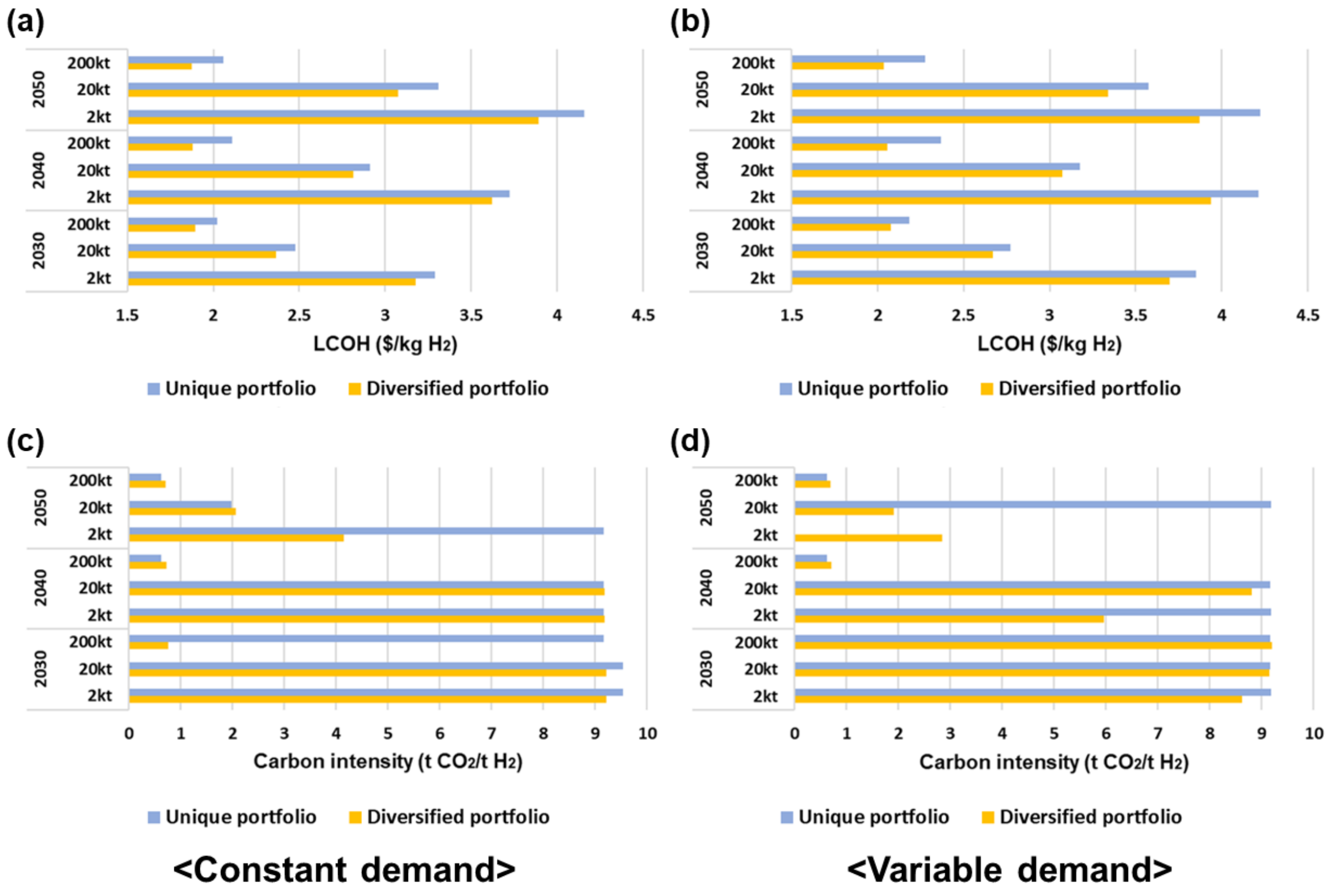

Fig. 7. No low C electricity option case: LCOH for Unique and Diversified portfolio: (a) Constant demand, (b) Variable demand, Carbon intensity for Unique and Base cases: (c) Constant demand. (d) Variable demand (Blue bar represents the Unique portfolio case, and Orange bar denotes the diversified portfolio case).

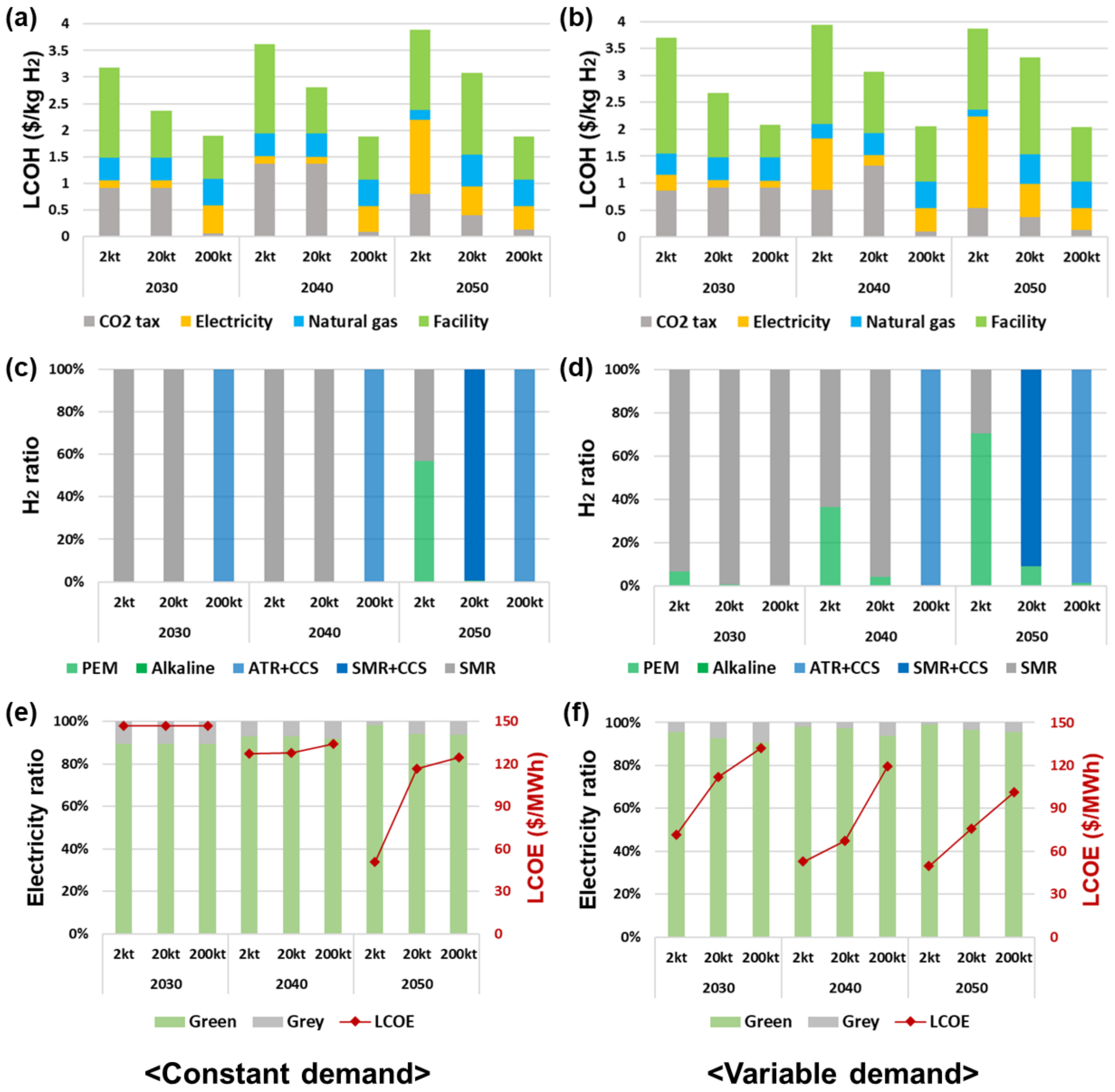

Fig. 8. No low C electricity option case: (a) LCOH breakdown for constant demand. (b) LCOH breakdown for variable demand. (c) Hydrogen production technology ratio for constant demand. (d) Hydrogen production technology ratio for variable demand. (e) Electricity ratio and LCOE for constant demand. (f) Electricity ratio and LCOE for variable demand.

### Expensive NG case: Influence of higher price of natural gas

This analysis examines the dynamics within the hydrogen and electricity supply chains under the condition where the price of NG triples from the base case, increasing from \$3.5/MMBtu to \$10.5/MMBtu. Fig. 9 presents the LCOH and carbon intensity for both the Unique case and the Base case while Fig. 10 illustrated the optimal result for the base case.

Key findings in LCOH perspective in Fig. 9:

- For constant demand, the LCOH for a diversified portfolio decreased by an average of 2% compared to a unique portfolio. The improvements ranged up to \$0.25/kg $H_2$ (equates to 8%).
- For variable demand, the LCOH reduction is more pronounced, around 4.6%. The improvements ranged up to \$0.65/kg $H_2$ (equates to 13%).
- This scenario highlights the growing importance of diversifying portfolios, especially as the price of NG increases and when there is variability in hydrogen demand.

Key findings in carbon intensity perspective in Fig. 9:

- For constant demand, the majority of carbon intensity reductions were observed, with an average decrease of 0.9 t $CO_2$/t $H_2$, equivalent to about 13%.
- For variable demand, there is no specific tendency, ranged from to 1.2 t $CO_2$/t $H_2$ reduction to 3.9 t $CO_2$/t $H_2$ increased. There is an significant increase in carbon intensity at 2 kt in 2040 and at 2 kt and 20 kt in 2050.
- Higher preference for green hydrogen in the unique portfolio, driven by the volatility of hydrogen demand and high natural gas prices. However, in a diversified portfolio, it may more profitable to integrate both SMR and green hydrogen.

Results are illustrated in Fig. 10 and notable observations are summarized below:

- There is a significant rise in the LCOH by over \$1/kg $H_2$. The proportion of NG costs within the LCOH markedly increases compared to the Base case.
- The reliance on blue hydrogen, which typically consumes more NG, notably decreases. This shift results in a concurrent reduction in the dependence on low C electricity.
- In the 2030 base scenario, substantial hydrogen demand predominantly utilized ATR+CCS-based blue hydrogen. With the price increase, the preference shifts to SMR-based grey hydrogen, which requires less NG.
- By 2050, for medium hydrogen demand scenarios, a mix of green and grey hydrogen becomes more favorable than SMR with CCS, marking a significant shift in the optimal supply chain strategy. Predominantly, green hydrogen production employs PEM electrolysis rather than alkaline electrolysis, due to its operational flexibility.
- Intriguingly, previous scenarios showed a correlation between an increased share of green hydrogen and a higher share of green electricity, leading to lower electricity costs. However, with the NG price hike, while the share of green hydrogen decreases from small to medium-scale demand by 2050, the share of green electricity actually rises, contributing to a decrease in electricity production costs. This is attributed to the fact that green hydrogen, which uses significantly more electricity per ton compared to SMRs, represents a large portion of the total electricity demand. The increasing role of grey hydrogen, which fulfills hydrogen needs with lower electricity usage, also helps minimize the need to meet operational thresholds, thus reducing reliance on grey and low C electricity.

- Despite the substantial increase in NG prices, ATR with CCS-based blue hydrogen remains the preferred technology for large-scale hydrogen demand in 2040 and 2050, maintaining an economic advantage of over \$1/kg $H_2$ compared to smaller demand scenarios.

This scenario underscores the substantial influence of NG prices on hydrogen production technology preferences and the broader implications for electricity consumption and cost within the hydrogen supply chain. The shift from blue to grey hydrogen, alongside an increased reliance on green electricity, illustrates how changes in fuel costs can reshape energy strategies, emphasizing the need for flexible and diverse energy solutions in response to volatile market conditions.

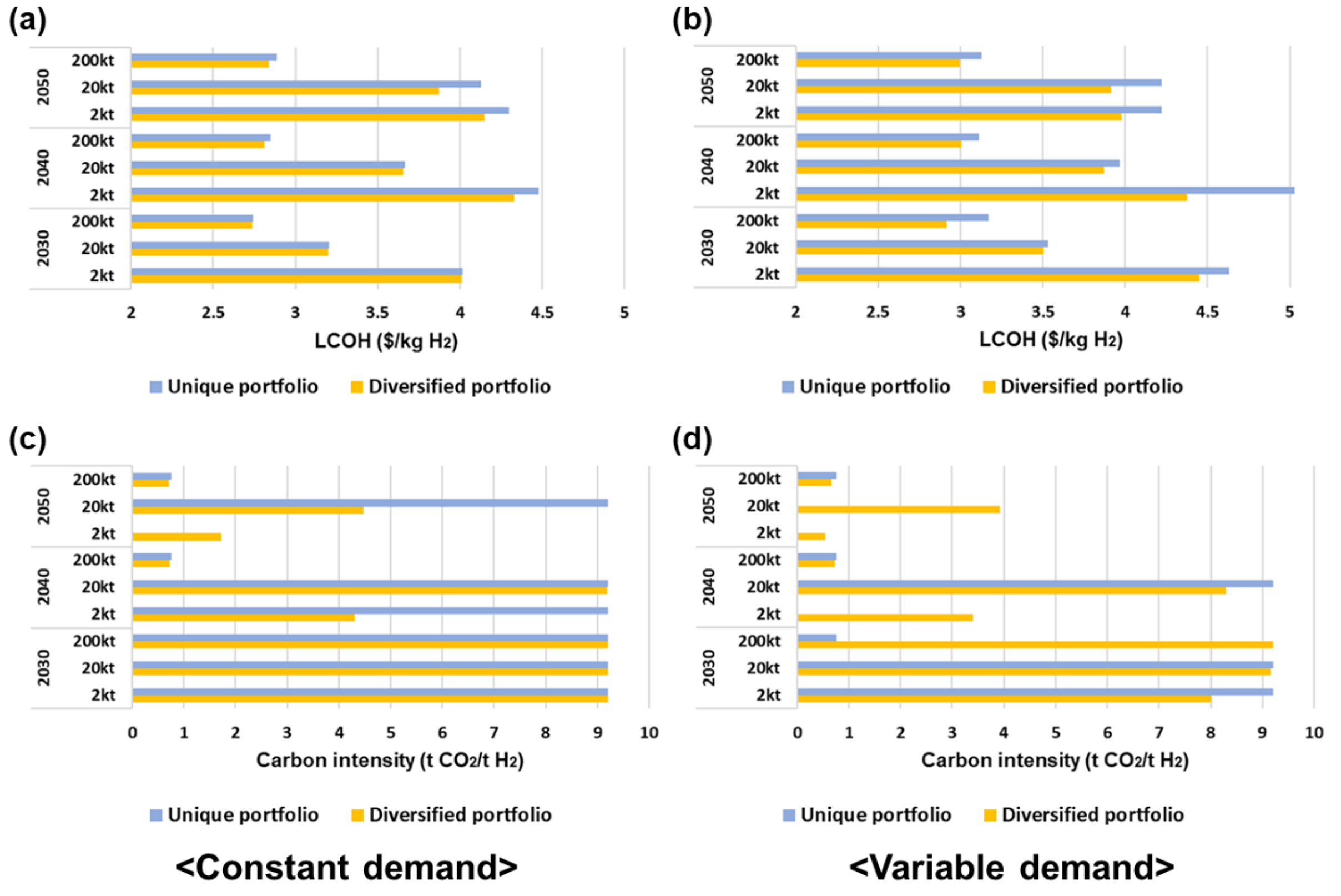

Fig. 9. Expensive NG case: LCOH for Unique and Diversified portfolio: (a) Constant demand, (b) Variable demand, Carbon intensity for Unique and Base cases: (c) Constant demand. (d) Variable demand (Blue bar represents the Unique portfolio case, and Orange bar denotes the diversified portfolio case).

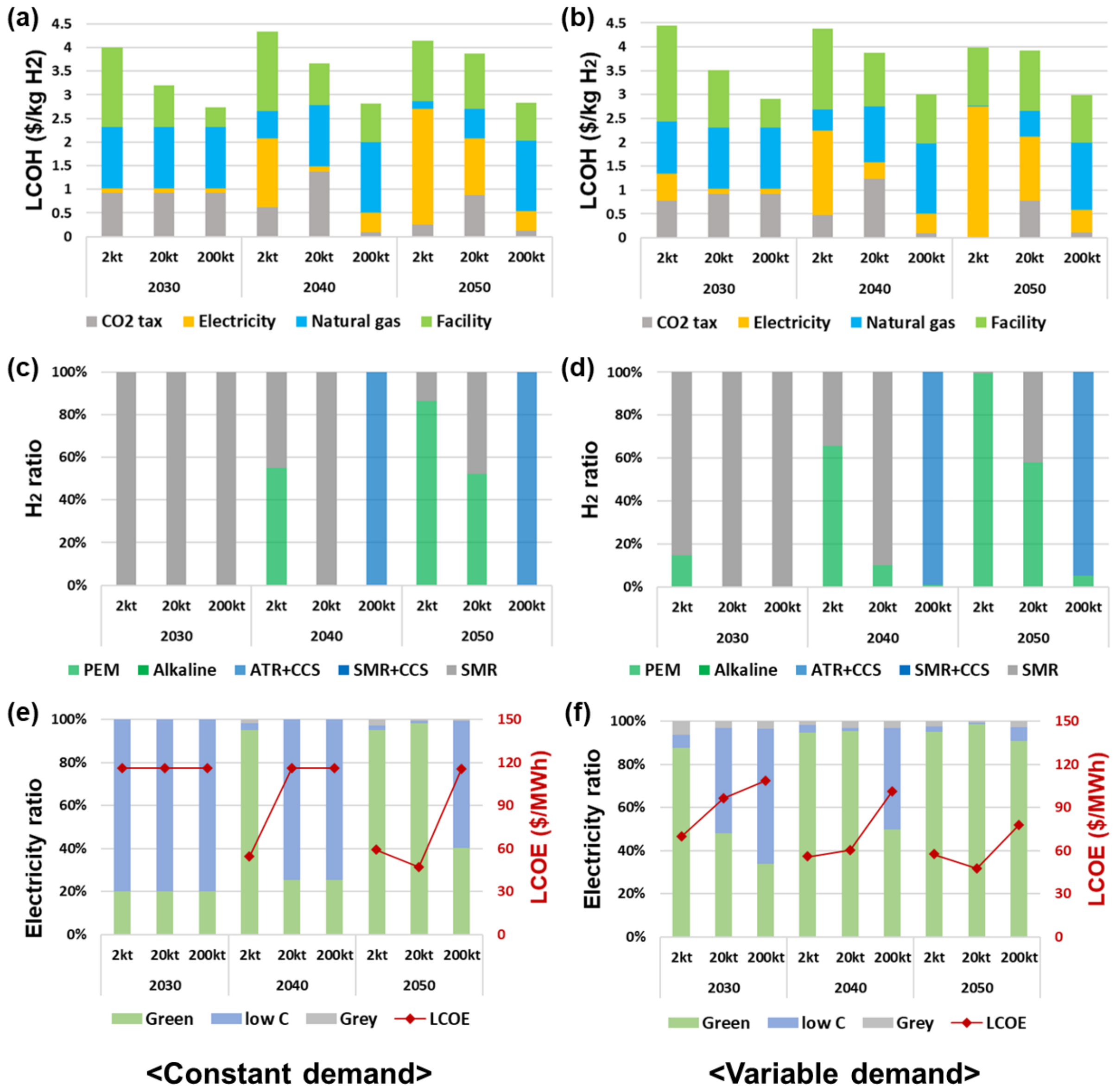

Fig. 10. Expensive NG case: (a) LCOH breakdown for constant demand. (b) LCOH breakdown for variable demand. (c) Hydrogen production technology ratio for constant demand. (d) Hydrogen production technology ratio for variable demand. (e) Electricity ratio and LCOE for constant demand. (f) Electricity ratio and LCOE for variable demand.

**Carbon tax sensitivity analysis in 2050: Influence of carbon tax in 2050**

This analysis explores how different carbon tax scenarios, specifically at \$150/t $CO_2$ and \$250/t $CO_2$, influence the hydrogen and electricity supply chains. These scenarios are based on a projected baseline carbon tax of \$200/t $CO_2$ for 2050. Results are presented in Fig. 11. The following points are noteworthy:

- At lower $CO_2$ tax levels, grey hydrogen has a significant presence in the LCOH, indicating a reliance on this variant. As the $CO_2$ tax increases, the contribution of grey hydrogen to the LCOH diminishes, transitioning towards more blue and green hydrogen. This shift is accompanied by a higher share of electricity and facility-related costs.
- At low hydrogen demand levels, the proportion of green hydrogen increases with rising $CO_2$ tax rates. This trend also sees an escalation in the usage of low C electricity for electricity production, especially for constant hydrogen demand scenarios where there is a marked preference for low C electricity. Industries with steady hydrogen needs tend to increase their use of low C electricity, utilizing alkaline electrolysis to reduce the costs of green hydrogen production, despite its potentially higher electricity production costs driven by its higher minimum operating threshold.
- Variable hydrogen demand introduces some utilization of alkaline electrolysis, though to a lesser extent compared to constant demand scenarios. The requirement for greater flexibility to manage demand volatility favors PEM electrolysis, which excels in operational flexibility. This leads to a diminished reliance on low C electricity in commercial settings, reinforcing the preference for PEM over alkaline electrolysis despite the cost advantages of the latter.
- For medium-scale hydrogen demand, the optimal hydrogen supply chain involves a blend of SMR-based grey hydrogen, which minimizes electricity consumption, and PEM electrolysis-based green hydrogen under lower $CO_2$ tax scenarios. Here, the share of green electricity is at its peak, minimizing the unit cost of electricity production. As $CO_2$ taxes increase, there is a slight increase in the share of green hydrogen, but the transition to blue hydrogen sourced from SMR with CCS reduces the share of green electricity, as SMR with CCS consumes significantly more electricity than standard SMR, thereby increasing the unit cost of electricity production. However, with even higher $CO_2$ tax levels, a marginal increase in green hydrogen's share causes a rise in green electricity utilization, lowering the unit cost of electricity production compared to the base case, despite the higher $CO_2$ tax.

Across other scenarios, the hydrogen and electricity supply chains are relatively stable under varying $CO_2$ tax rates. However, increases in $CO_2$ tax levels lead to slight rises ($2-3/MWh) in the unit cost of electricity production due to $CO_2$ emissions from low C electricity.

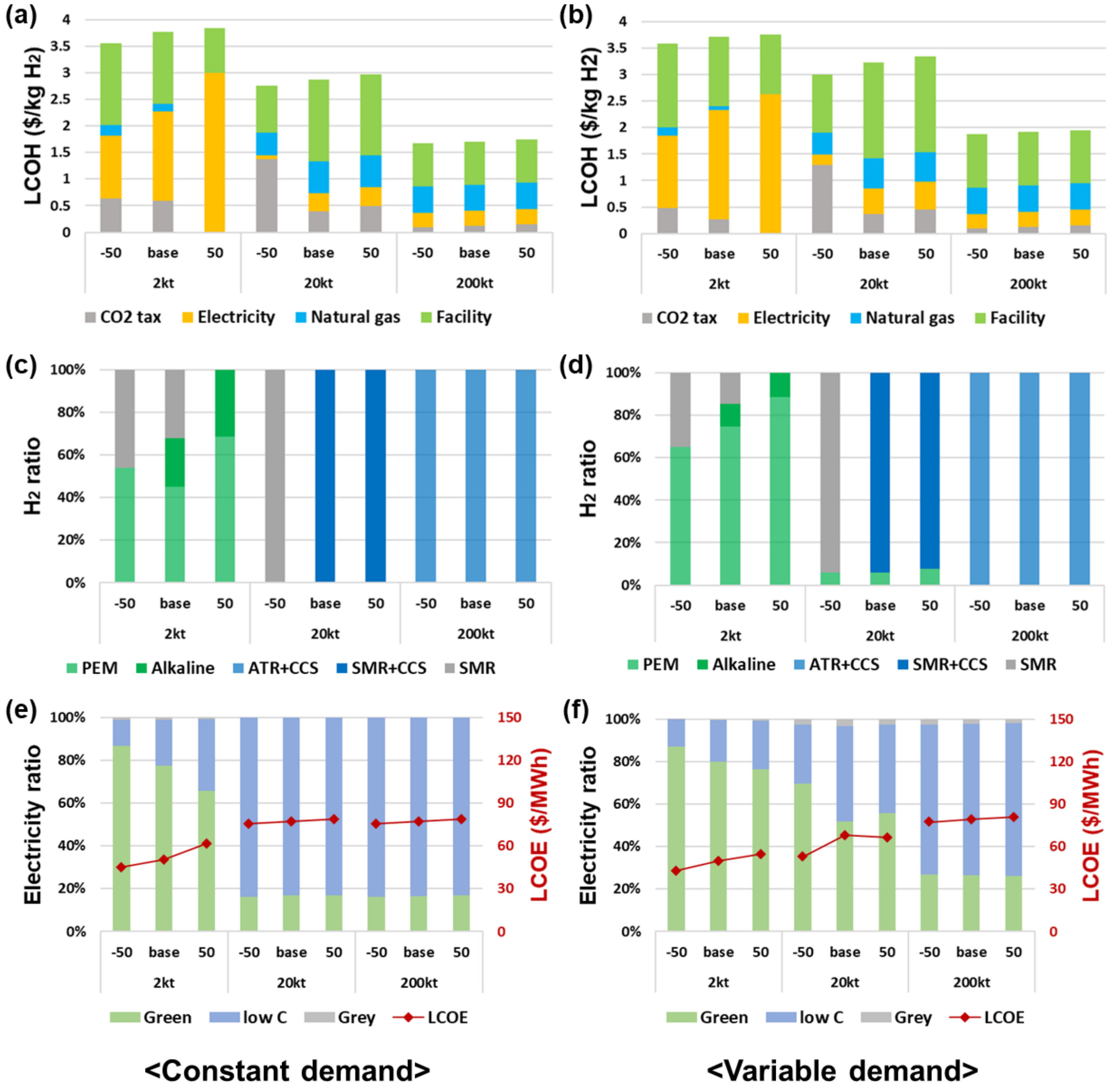

Fig. 11. Carbon tax sensitivity analysis in 2050: (a) LCOH breakdown for constant demand. (b) LCOH breakdown for variable demand. (c) Hydrogen production technology ratio for constant demand. (d) Hydrogen production technology ratio for variable demand. (e) Electricity ratio and LCOE for constant demand. (f) Electricity ratio and LCOE for variable demand.

## Discussion

Our study delves into the co-optimization of hydrogen and electricity supply chains, highlighting the necessity of addressing various scales and types of hydrogen demand. We advocate for a diversified approach to configuring supply chains, leveraging a mix of technologies instead of relying solely on a single method for hydrogen and electricity production. This strategy underscores the intricate relationship between hydrogen and electricity supply chains, revealing the nuanced dynamics that inform optimal configurations. The key findings are summarized:

- At smaller scales of hydrogen demand, our findings indicate a preference for Steam Methane Reforming (SMR) technology for constant demand scenarios. However, for fluctuating hydrogen demand, Proton Exchange Membrane (PEM) electrolysis is favored due to its lower operational constraints. Additionally, as the share of green hydrogen increases, it results in a higher proportion of green electricity, thereby reducing the Levelized Cost of Energy (LCOE) and carbon intensity.
- In scenarios of increased hydrogen demand, we observe a preference for blue hydrogen, enabled by the scalability of Auto-Thermal Reforming (ATR) with Carbon Capture and Storage (CCS) technology, which benefits from economies of scale. Conversely, in commercial settings characterized by variable hydrogen demand, there is a greater reliance on green hydrogen. The operational flexibility and cost-effectiveness of green electricity are crucial in these cases, shaping the supply chain configurations.
- The lack of low-carbon electricity presents significant challenges, raising electricity production costs and necessitating a shift toward grey hydrogen to minimize electricity consumption. The viability of alkaline electrolysis in this context is limited by higher electricity costs. Our findings highlight the importance of a diversified portfolio, especially in the absence of low-carbon electricity, and underscore the crucial role of low-carbon electricity in facilitating a cleaner energy transition.
- External factors, such as fluctuations in natural gas prices and varying CO2 tax rates, significantly impact supply chain dynamics. A substantial increase in natural gas prices enhances the benefits of a diversified portfolio, prompting a shift from blue to grey hydrogen, reflecting the supply chain's sensitivity to market dynamics. Similarly, rising CO2 taxes drive the adoption of cleaner hydrogen production methods, altering the cost structure of hydrogen production.

In conclusion, our comprehensive study emphasizes the complex nature of hydrogen and electricity supply chains. It underscores the importance of configurations that consider demand characteristics, technological factors, and external market forces. Our insights provide valuable guidance for policymakers, industry stakeholders, and researchers aiming to navigate the transition towards sustainable energy systems.

## Method

### Solar irradiance scenarios

We employed hourly solar Global Horizontal Irradiance (GHI) data from 2019 to 2021, gathered near Mount Signal Solar Power plant (latitude: 32.8, longitude: -115.8). The data was sourced from the National Solar Radiation Database (NSRDB) [26].

### California hydrogen demand scenarios

We conduct a detailed analysis of how different types of hydrogen demand—constant and variable—and scales ranging from 2 kt to 200 kt per year influence the economics and optimal configuration of hydrogen and electricity supply chains. The current production levels of California's SMR plants, approximately 1.8 Mt per year, anticipate a significant increase in variable hydrogen demand, projected to rise by about 1.7 Mt per year by 2050. In anticipation of this growth, policies are being carefully developed to ensure that both existing and future hydrogen production meets environmental standards, demonstrating California's commitment to sustainable energy practices. Additionally, the variable demand profile is based on the hourly commercial electricity demand data for 2019-2021, as reported by the Energy Information Administration (EIA).

### Mathematical programming

Two-stage Stochastic Programming (2SSP) is utilized to determine the optimal configuration and facilitate operational decisions that consider uncertainties. The 2SSP framework requires decision-makers to make two distinct sets of decisions. First-stage decisions are made before the uncertainties are known, primarily concerning the design aspects such as capacity planning and infrastructure investments. The second-stage decisions are made after the uncertainties, such as solar irradiance and hydrogen demand, are realized, allowing for operational adjustments based on the initial decisions. These adjustments are customized for each facility's specific circumstances. Detailed formulations and explanations of this approach are available in the Supplementary Materials, and the corresponding code can be accessed via a link on GitHub.

### Acknowledgements

This work was supported by the KAIST-Aramco $CO_2$ management center.

### Competing interests

The authors declare no competing interests.

# Multi-technology co-optimizing approach for sustainable hydrogen and electricity supply chains considering variability and demand scale

## Supporting information

Sunwoo Kim[1,2], Joungho Park[3], and Jay H. Lee[1*]

[1] Mork Family Department of Chemical Engineering and Material Sciences, University of Southern California, 925 Bloom Walk, Los Angeles, CA 90089, USA

[2] Department of Chemical and Biomolecular Engineering, Korea Advanced Institute of Science and Technology (KAIST), 291, Daehak-ro, Yuseong-gu, Daejeon, 34141, Republic of Korea

[3] Energy AI & Computational Science Laboratory, Korea Institute of Energy Research, 71-2 Jang-dong, Yuseong-gu, Daejeon, 305-343, Republic of Korea

[*] Corresponding author's E-mail: jlee4140@usc.edu

---

Table of contents

**A. Models used**

A. 1. Battery

This study employs a 4-hour battery system, designed to complete a full charge and discharge cycle within a four-hour period, as detailed in eqn (A3) and (A4). The dynamics of energy storage within the battery ($ESS_{sc,t}$, unit: MWh) are governed by various factors such as the self-discharging rates ($\sigma^B$), the charging ($ch_{sc,t}$, unit: MW) and discharging ($dch_{sc,t}$, unit: MW) rates, the overall battery efficiency ($\eta^B$), and DC/DC converter efficiency ($\eta^{DC/DC}$, 98%), as elaborated in eqn (A5). The self-discharge rate is assumed to be 0.0083% per hour and the battery efficiency is set at 95%. Furthermore, it is stipulated that the battery's storage level must be maintained within 15% to 95% of its rated capacity, as expressed in eqn (A6). Lifespan is assumed to be 10 years and yearly fixed O&M cost is assumed to be 2.5% of installation cost [1].

$$0 \le ch_{sc,t} \le 0.25 \times X_B \tag{A3}$$

$$0 \le dch_{sc,t} \le 0.25 \times X_B \tag{A4}$$

$$ESS_{sc,t} = \left(1 - \sigma^B\right) \times ESS_{sc,t-1} + \left(\eta^{DC/DC} \times \eta^B \times ch_{sc,t} - \frac{1}{\eta^{DC/DC} \times \eta^B} \times dch_{sc,t}\right) \times \Delta t \tag{A5}$$

$$0.15 \times X_B \le ESS_{sc,t} \le 0.95 \times X_B \tag{A6}$$

A. 2. Solar PV panel

The power output from a solar photovoltaic (PV) panel is defined by the following equation [2, 3]:

$$P_{sc,t}^{P} = \eta^{DC/DC} \times X_P \times \left[1 + \gamma \times \left(T^{sc,t} + GHI^{sc,t} \times \frac{NOCT - 20}{800} - T^{ref}\right)\right] \times \frac{GHI^{sc,t}}{1000} \tag{A2}$$

In this equation, $P_{sc,t}^{P}$ represents the power generated from the PV panel, $\eta^C$ denotes the efficiency of the appropriate converter, and $X_P$ is the PV panel's capacity. $\gamma$ (-0.0037 $K^{-1}$), is the temperature coefficient, affecting the panel's performance relative to the cell temperature ($T^{sc,t}$). $GHI^{sc,t}$ stands for the global horizontal solar irradiance (GHI) at a specific scenario and time, the normal operating cell temperature ($NOCT$) is set at 45 ℃, and $T^{ref}$ (298.15 K) represents the reference ambient temperature. The assumed lifetime of the PV system is 35 years and yearly fixed O&M cost is estimated to be 2% of the installation cost.

A. 3. NG power plant with carbon capture

This study investigates the use of NG power plants equipped with CC as a viable option for generating low C electricity. The NG power plant produces electricity and heat, with its flue gases directed into a carbon capture system [4, 5]. Within this process, $CO_2$ is extracted from the flue gas stream using an amine-based solvent which, while reducing energy efficiency, significantly lowers the carbon intensity to 0.038 t $CO_2$/MWh. The minimum operating rate is assumed to be 60% of their rated capacity ($X_{NGCC}$, unit: MW). The investment cost for this technology is estimated at 3 MM$/MW, and the variable costs amount to $5.6/MWh. Additionally, 7.15 MMBtu of NG is required to generate 1 MWh of power. The assumed lifespan of these plants is 25 years, with yearly fixed operations and maintenance (O&M) costs calculated at 3% of the installation cost [6].

$$0.6 \times X_{NGCC} \le NGCC_{sc,t} \le X_{NGCC} \quad \text{(A7)}$$

In above equation, $NGCC_{sc,t}$ indicates the power output rate (in MW) generated by the NG power plant equipped with CC technology.

A. 4. Main grid

In this study, we consider an additional electricity supply option—the conventional main grid, which typically has a higher carbon intensity compared to the aforementioned low C options. In California, the carbon intensity of the main grid electricity is estimated at 0.376 t $CO_2$/MWh, with the commercial electricity price assumed to be $227/MWh [7, 8]. Looking ahead, it is expected that the primary electricity grid will become the most expensive option due to escalating $CO_2$ taxes, relative to green or low C alternatives. Our research highlights the economic benefits of sourcing locally generated green or low-carbon electricity over purchasing from the main grid, particularly within the context of California. Furthermore, the anticipated increase in preference for low-carbon or green electricity, spurred by the implementation of $CO_2$ taxes, indicates a potential broader adoption beyond California. This trend underscores a consistent transition towards more sustainable energy practices across different regions.

A. 5. Green hydrogen technology (PEM, Alkaline electrolysis)

This study incorporates the use of water electrolyzers to split water into hydrogen and oxygen gases. The PEM electrolyzer features a solid polymer membrane that conducts protons from the anode to the cathode, effectively preventing the mixing of the two gases. In contrast, the alkaline electrolyzer operates with two electrodes submerged in a liquid alkaline electrolyte.

The two technologies are distinguished by their ability to produce high-purity hydrogen gas and their capacity to operate efficiently at low minimum loads ($min^W$)—typically 5% for the PEM electrolyzer and 20% for alkaline electrolyzers[9, 10].

Furthermore, the assumed energy efficiency ($\eta^W$) for the PEM electrolyzer is 48 kWh per kilogram of hydrogen, and for the Alkaline electrolyzer, it is 50 kWh per kilogram of hydrogen.

These efficiencies reflect the power-to-hydrogen conversion rates ($p2h_{sc,t}$) and the capacities ($X_W$) of the electrolyzers, both expressed in MW. Detailed operational parameters and constraints for these water electrolyzers are elaborated in the following equation (A7).

$$\min^W \times X_W \leq p2h_{sc,t} \leq X_W \tag{A7}$$

Both types of electrolyzers are assumed to have a lifespan of 10 years, with yearly fixed O&M costs estimated at 2% of the installation cost.

A. 6. NG reforming hydrogen technology (ATR with CC, SMR with CC, and SMR)

This study evaluates three NG reforming technologies for producing blue and grey hydrogen. SMR is a chemical process that combines steam and methane to produce hydrogen through a catalytic reaction, commonly employed in industrial hydrogen production[11]. ATR is a variation of SMR that introduces both steam and oxygen, facilitating simultaneous reforming and partial oxidation reactions that enhance hydrogen production efficiency [12]. When integrated with CC technology, both types of reformers can significantly reduce $CO_2$ emissions from hydrogen production. SMR with CC involves capturing $CO_2$ emitted during the methane reforming process, whereas ATR with CC can achieve potentially higher efficiency by capturing $CO_2$ from both the reforming and partial oxidation steps [12, 13]. These differences allow the reformers to exhibit varying economies of scale, as depicted in Figure A1. Additionally, the energy consumption rates and $CO_2$ emission rates for these technologies are detailed in Table A.1, providing a comprehensive overview of their environmental impact and operational efficiency.

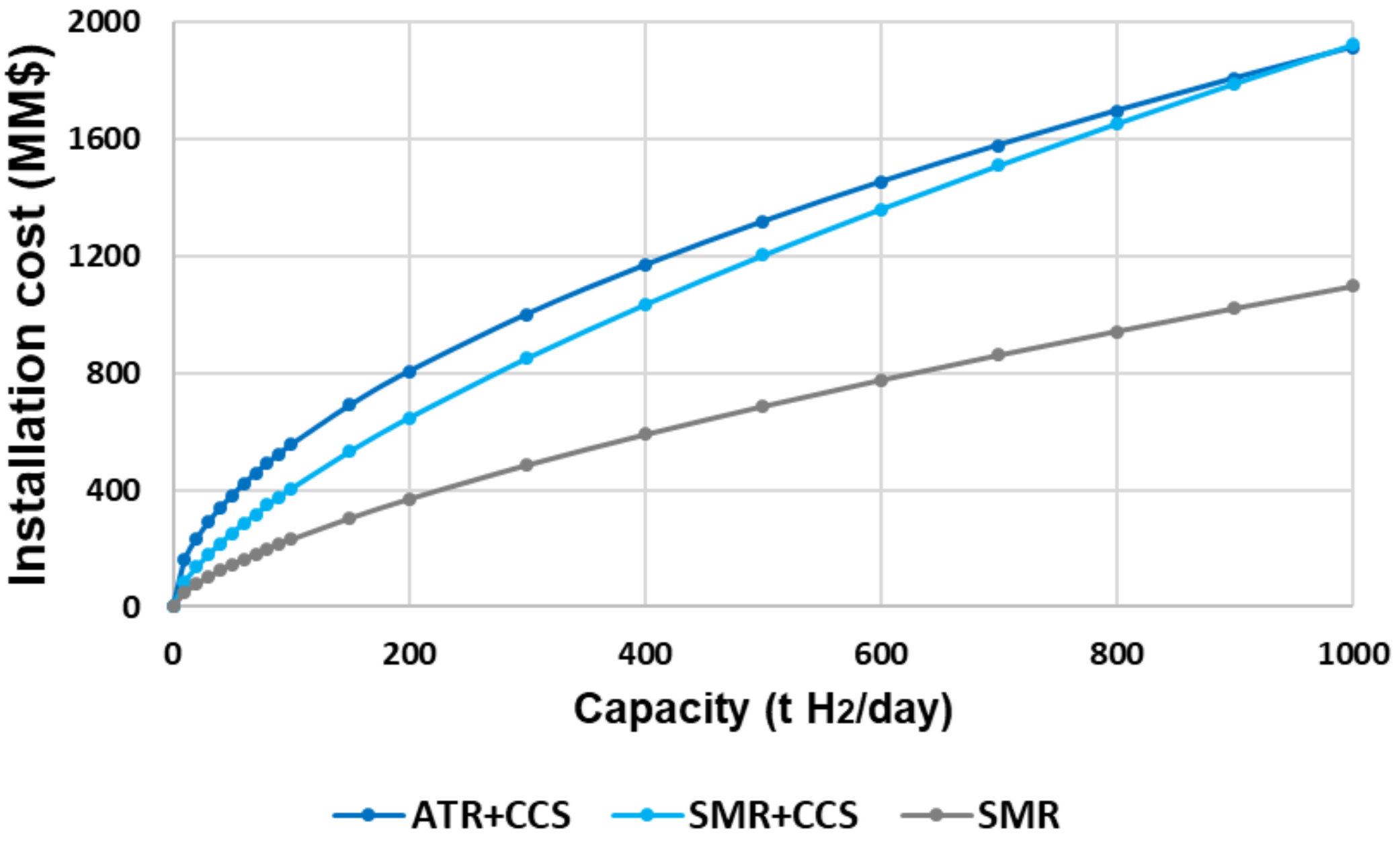

Fig. A 1. Nonlinear installation cost function for each reformers [13].

Table. A. 1. Summary of NG and electricity consumption rates, and $CO_2$ emission rates for each reformer [13].

| | ATR with CC | SMR with CC | SMR |
|---|---|---|---|
| **NG consumption rate** | 142 MMBtu/t $H_2$ | 171 MMBtu/t $H_2$ | 123 MMBtu/t $H_2$ |
| **Electricity consumption rate** | 3.6 MWh/t $H_2$ | 4.4 MWh/t $H_2$ | 0.96 MWh/t $H_2$ |
| **$CO_2$ emission rate** | 0.62 t $CO_2$/t $H_2$ | 1.98 t $CO_2$/t $H_2$ | 9.17 t $CO_2$/t $H_2$ |

The nonlinear installation cost is approximated as piecewise linear function to reduce the complexity of the optimization programming by following equations:

$$IC_{NR} = IC_{NR}^{1} \times x_1 + IC_{NR}^{2} \times x_2 + ... + IC_{NR}^{k} \times x_k \quad \text{where} \quad \sum_k x_k \leq 1, \; 0 \leq x_k \leq 1 \tag{A1}$$

$$X_{NR} = X_{NR}^{1} \times x_1 + X_{NR}^{2} \times x_2 + ... + X_{NR}^{k} \times x_k \tag{A2}$$

$$\sum_k y_k \leq 2 \quad \text{where} \quad x_k \leq y_k, \; y_k \in \{0,1\} \tag{A3}$$

$$y_k + y_j \leq 1 \text{ where } k, j \text{ is not adjacent} \tag{A4}$$

For all reformers, the minimum operating load is set at 60%, and the ramp rate—indicating the change in operating rate ($n2h_{sc,t}$, unit: MW) within an hour —is assumed to be 20% of their rated capacity ($X_{NR}$, unit: MW). The operational parameters and constraints of the reformers are elaborated in the subsequent equations:

$$0.6\times X_{NR}\leq n2h_{sc,t}\leq X_{NR} \tag{A7}$$

$$-0.2\times X_{NR}\leq n2h_{sc,t}-n2h_{sc,t-1}\leq 0.2\times X_{NR} \tag{A8}$$

The lifespan of all reformers is assumed to be 25 years, and the yearly fixed O&M costs are assumed to be 4% of the installation cost.

A. 7. Gaseous hydrogen tank

This research utilizes a gaseous hydrogen tank (HT) system, designed to store hydrogen under high pressure to efficiently accommodate large volumes of hydrogen gas[14]. The volume of hydrogen stored in the HT ($HT_{sc,t}$, unit: ton of hydrogen) is determined by various factors, including the self-leakage rate ($\sigma^{HT}$), and the rate of hydrogen production from the electrolyzer, with $D_{sc,t}$ representing the demand of hydrogen that needs to be met. Consequently, the balance equation for the HT is adjusted to reflect this demand, as detailed in equation (A10). The self-leakage rate of the HT is assumed to be 0.0104% per hour, which accounts for minimal hydrogen loss over time due to imperfections in the containment system of the tank. Additionally, it is crucial that the HT's storage level does not surpass its designed capacity to ensure safety and maintain the integrity of the storage system, as indicated in eqn (A11). This constraint is essential for preventing overpressure scenarios and ensuring the operational reliability of the hydrogen storage system. The lifespan is assumed to be 25 years and yearly fixed O&M cost is assumed to be 1% of the installation cost.

$$HT_{sc,t}=\left(1-\sigma^{HT}\right)\times HT_{sc,t-1}+\left(\frac{1}{\eta^{W}}\times p2h_{sc,t}+\frac{1}{\eta^{NR}}\times n2h_{sc,t}\right)\times\Delta t-D_{sc,t} \tag{A10}$$

$$0\leq HT_{sc,t}\leq X_{HT} \tag{A11}$$

## B. Energy balance equations

The electricity balance equation for the DC bus and AC bus are captured in the equations below:

$$P^{P}_{sc,t} - ch_{sc,t} + dch_{sc,t} - D2A_{sc,t} + \eta^{AC/DC} \times A2D_{sc,t} - out_{sc,t} - p2h_{sc,t} = 0 \tag{B1}$$

$$NGCC_{sc,t} + MG_{sc,t} + \eta^{AC/DC} \times D2A_{sc,t} - A2D_{sc,t} - n2h_{sc,t} = 0 \tag{B2}$$

In these equations, the $D2A_{sc,t}$ indicates the power rate from the DC bus to the AC bus, $A2D_{sc,t}$ denotes the power transform the AC bus to the DC bus, and $out_{sc,t}$ represents the curtailment. The efficiency of the converter ($\eta^{AC/DC}$) is incorporated into these equations to account for energy losses during conversion, and it is assumed to be 95%.

## C. Objective function

The objective function represents the total annualized cost as presented in the equations below:

$$\sum_{i}\left[\left(crf^{i} + omf^{i}\right) \times IC_{i}\right] + \frac{1}{N^{sc}} \times \sum_{sc}\sum_{t}\sum_{y}\left[y_{sc,t} \times \left(ng^{y} \times c^{NG} + co_{2}^{\,y} \times c^{CO_2} + c^{Variable}\right)\right] \tag{C1}$$

$$crf^{i} = \frac{r \times (1+r)^{L^{i}}}{(1+r)^{L^{i}} - 1} \tag{C2}$$

In these equations, $crf^{i}$ denotes the capital recovery factor, $omf^{i}$ represents the fraction of the installation cost ($IC^{i}$) allocated for yearly operating and maintenance costs of facility $i$. $N^{sc}$ represents the total number of scenarios, indexed by the subscript $sc$. $y_{sc,t}$ indicates the operational decision variables, including $NGCC_{sc,t}$, $n2h_{sc,t}$. The costs integrated into the objective function include the natural gas consumption rate ($ng^{y}$) multiplied by the natural gas cost ($c^{NG}$), the $CO_2$ emission rate ($co_2{}^{y}$) multiplied by the $CO_2$ tax ($c^{CO_2}$), and the variable cost ($c^{Variable}$). In this model, the interest rate ($r$) is assumed to be 7%, and $L^{i}$ represents the lifespan of facility $i$.

## D. Ideal (Unrealistic) case: Constant renewable supply and hydrogen demand

In this scenario, constant availability of renewable energy and steady hydrogen demand are assumed to explore the economic feasibility of different hydrogen production technologies and scales. The results are summarized in Fig. D 1. Notable findings can be summarized as below:

- Fig. D 1 (a): By 2030, the cost of electricity stands at approximately \$2.4/kg $H_2$ for a hydrogen demand of 2 kt/year, which significantly reduces to about \$1.5/kg $H_2$ for a demand of 200 kt/year. For smaller scales (2 kt per year), the unit cost of hydrogen decreases notably to around \$1.8/kg $H_2$ from 2030 to 2050 due to advancements in green hydrogen and electricity technologies. However, for larger scales (200 kt per year), the production costs experience a slight increase, driven by rising annual $CO_2$ taxes. Green hydrogen predominantly incurs costs from electricity, while grey hydrogen sees significant impacts from $CO_2$ tax and facility costs, and blue hydrogen costs are heavily influenced by facility and natural gas expenses.
- Fig. D 1 (b): At the smaller demand scale, alkaline electrolysis is favored over PEM electrolysis due to its lower installation and operating costs, despite the slightly higher efficiency (~4%) of PEM. The cost differential between these technologies, which ranges from 30-40%, positions alkaline electrolysis as the more cost-effective choice. Conversely, for the larger demand scale, ATR+CCS is preferred over SMR+CCS due to ATR's superior carbon capture efficiency. The higher initial investment required for ATR+CCS is offset by its lower natural gas and electricity use, coupled with a higher carbon capture rate.
- Fig. D 1 (c): As technology advances, there is a noticeable shift in preference at the medium-scale demand level (20 kt per year). The preference transitions from grey hydrogen produced by SMRs in 2030 to blue hydrogen from ATR with CCS by 2040, ultimately shifting to green hydrogen produced by alkaline electrolysis by 2050.
- General Observations: In all scenarios, green electricity sourced from PV panels is used, with the unit cost of electricity estimated at approximately \$40/MWh in 2030, declining to around \$30/MWh by 2050 due to technological advancements. The use of DC electricity for green hydrogen results in a 5% lower LCOE compared to blue and grey hydrogen technologies that utilize AC electricity.

It is critical to note that this analysis omits considerations for energy storage solutions like batteries or hydrogen tanks, which may overestimate the economic viability of solar power under continuous operation scenarios.

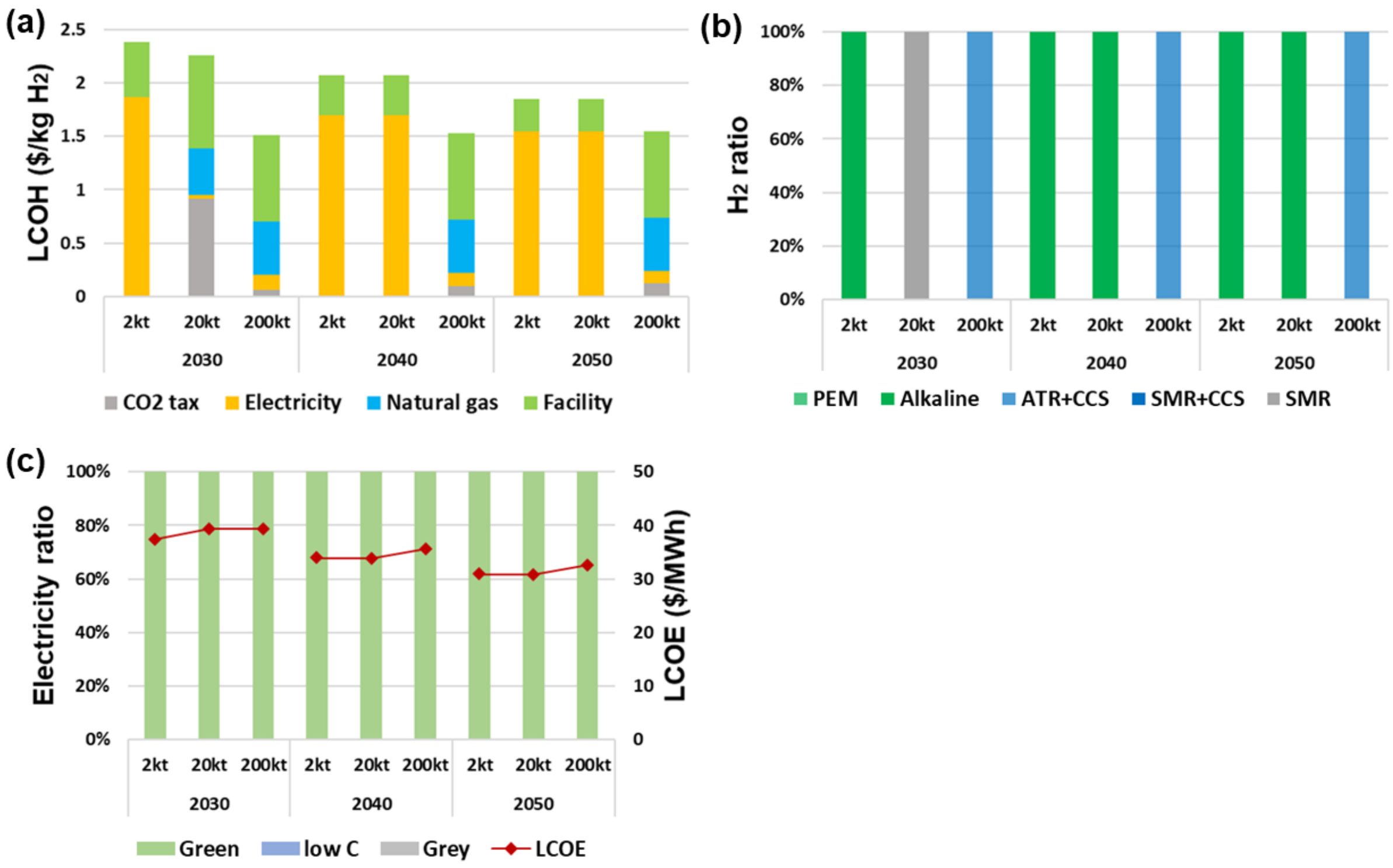

Fig. D 1. Ideal (Unrealistic) case: (a) LCOH breakdown: This graph displays the LCOH composition for $CO_2$ tax, electricity, NG, and facility costs across different hydrogen demand scales from 2030 to 2050. It illustrates how each cost component varies with changes in demand scale. (b) Hydrogen production technology ratio: This chart shows the proportion of hydrogen produced using different technologies—PEM electrolysis, alkaline electrolysis, ATR+CCS, SMR+CCS, and SMR—and how these proportions shift with varying demand scales over the period from 2030 to 2050. (c) Electricity consumption and LCOE: This graph presents the ratio of electricity consumption and the LCOE for varying demand scales from 2030 to 2050.

## E. Optimal configuration result tables

The detailed design capacity of each facilities and the emission rate of $CO_2$ and consumption rate of NG is summarized in below tables for corresponding Cases.

Table E. 1. Unrealistic case: summary of optimal configuration, CO2 emission rate, NG consumption rate.

| | | PV (MW) | B (MWh) | PEM (MW) | Alkaline (MW) | H tank (t $H_2$) | NGCC (MW) | ATR (t $H_2$/day) | SMRCC (t $H_2$/day) | SMR (t $H_2$/day) | $CO_2$ (kt/year) | NG (TBtu/year) |
|---|---|---|---|---|---|---|---|---|---|---|---|---|
| 2030 | 2 kt | 51.28 | 0 | 0 | 11.42 | 0 | 0 | 0 | 0 | 0 | 0 | 0 |
| | 20 kt | 10.36 | 0 | 0 | 0 | 0 | 0 | 0 | 0 | 54.79 | 183.4 | 2.460 |
| | 200 kt | 387.55 | 0 | 0 | 0 | 0 | 0 | 547.95 | 0 | 0 | 124.0 | 28.400 |
| 2040 | 2 kt | 51.28 | 0 | 0 | 11.42 | 0 | 0 | 0 | 0 | 0 | 0 | 0 |
| | 20 kt | 512.77 | 0 | 0 | 114.16 | 0 | 0 | 0 | 0 | 0 | 0 | 0 |
| | 200 kt | 387.55 | 0 | 0 | 0 | 0 | 0 | 547.95 | 0 | 0 | 124.0 | 28.400 |
| 2050 | 2 kt | 51.28 | 0 | 0 | 11.42 | 0 | 0 | 0 | 0 | 0 | 0 | 0 |
| | 20 kt | 512.77 | 0 | 0 | 114.16 | 0 | 0 | 0 | 0 | 0 | 0 | 0 |
| | 200 kt | 387.55 | 0 | 0 | 0 | 0 | 0 | 547.95 | 0 | 0 | 124.0 | 28.400 |

Table E. 2. Unique case: summary of optimal configuration, $CO_2$ emission rate, NG consumption rate for constant $H_2$ demand.

| | | PV (MW) | B (MWh) | PEM (MW) | Alkaline (MW) | H tank (t $H_2$) | NGCC (MW) | ATR (t $H_2$/day) | SMRCC (t $H_2$/day) | SMR (t $H_2$/day) | $CO_2$ (kt/year) | NG (TBtu/year) |
|---|---|---|---|---|---|---|---|---|---|---|---|---|
| 2030 | 2 kt | 0 | 0 | 0 | 0 | 0 | 0.22 | 0 | 0 | 5.48 | 18.4 | 0.260 |
| | 20 kt | 0 | 0 | 0 | 0 | 0 | 2.19 | 0 | 0 | 54.79 | 184.1 | 2.597 |
| | 200 kt | 0 | 0 | 0 | 0 | 0 | 81.96 | 547.95 | 0 | 0 | 151.3 | 33.534 |
| 2040 | 2 kt | 0 | 0 | 0 | 0 | 0 | 0.22 | 0 | 0 | 5.48 | 18.4 | 0.260 |
| | 20 kt | 0 | 0 | 0 | 0 | 0 | 2.19 | 0 | 0 | 54.79 | 184.1 | 2.597 |
| | 200 kt | 0 | 0 | 0 | 0 | 0 | 81.96 | 547.95 | 0 | 0 | 151.3 | 33.534 |
| 2050 | 2 kt | 0 | 0 | 0 | 0 | 0 | 0.22 | 0 | 0 | 5.48 | 18.4 | 0.260 |
| | 20 kt | 0 | 0 | 0 | 0 | 0 | 10.05 | 0 | 54.79 | 0 | 42.9 | 4.049 |
| | 200 kt | 0 | 0 | 0 | 0 | 0 | 81.96 | 547.95 | 0 | 0 | 151.3 | 33.534 |

Table E. 3. Unique case: summary of optimal configuration, $CO_2$ emission rate, NG consumption rate for variable $H_2$ demand.

| | | PV (MW) | B (MWh) | PEM (MW) | Alkaline (MW) | H tank (t $H_2$) | NGCC (MW) | ATR (t $H_2$/day) | SMRCC (t $H_2$/day) | SMR (t $H_2$/day) | $CO_2$ (kt/year) | NG (TBtu/year) |
|---|---|---|---|---|---|---|---|---|---|---|---|---|
| 2030 | 2 kt | 0 | 0 | 0 | 0 | 14.51 | 0.27 | 0 | 0 | 6.66 | 18.4 | 0.260 |
| | 20 kt | 0 | 0 | 0 | 0 | 44.07 | 2.95 | 0 | 0 | 73.64 | 184.1 | 2.597 |
| | 200 kt | 0 | 0 | 0 | 0 | 395.38 | 111.05 | 742.41 | 0 | 0 | 151.3 | 33.536 |
| 2040 | 2 kt | 0 | 0 | 0 | 0 | 14.51 | 0.27 | 0 | 0 | 6.66 | 18.4 | 0.260 |
| | 20 kt | 0 | 0 | 0 | 0 | 46.52 | 2.93 | 0 | 0 | 73.36 | 184.1 | 2.597 |
| | 200 kt | 0 | 0 | 0 | 0 | 465.24 | 109.73 | 733.57 | 0 | 0 | 151.3 | 33.536 |
| 2050 | 2 kt | 70.37 | 29.90 | 30.25 | 0 | 76.49 | 0.00 | 0 | 0 | 0 | 0 | 0 |
| | 20 kt | 0 | 0 | 0 | 0 | 158.23 | 12.12 | 0 | 66.13 | 0 | 43.0 | 4.052 |
| | 200 kt | 0 | 0 | 0 | 0 | 868.38 | 105.15 | 702.98 | 0 | 0 | 151.3 | 33.536 |

Table E. 4. Unique and no low C case: summary of optimal configuration, $CO_2$ emission rate, NG consumption rate for constant $H_2$ demand.

| | | PV (MW) | B (MWh) | PEM (MW) | Alkaline (MW) | H tank (t $H_2$) | NGCC (MW) | ATR (t $H_2$/day) | SMRCC (t $H_2$/day) | SMR (t $H_2$/day) | $CO_2$ (kt/year) | NG (TBtu/year) |
|---|---|---|---|---|---|---|---|---|---|---|---|---|
| 2030 | 2 kt | 0 | 0 | 0 | 0 | 0 | 0 | 0 | 0 | 5.48 | 19.1 | 0.246 |
| | 20 kt | 0 | 0 | 0 | 0 | 0 | 0 | 0 | 0 | 54.79 | 190.6 | 2.460 |
| | 200 kt | 311.74 | 654.80 | 0 | 0 | 129.78 | 0 | 0 | 0 | 551.58 | 1834.1 | 24.602 |
| 2040 | 2 kt | 2.80 | 8.09 | 0 | 0 | 0.43 | 0 | 0 | 0 | 5.48 | 18.3 | 0.246 |
| | 20 kt | 28.33 | 79.28 | 0 | 0 | 5.23 | 0 | 0 | 0 | 54.84 | 183.4 | 2.460 |
| | 200 kt | 976.97 | 1820.96 | 0 | 0 | 617.51 | 0 | 580.43 | 0 | 0 | 124.0 | 28.407 |
| 2050 | 2 kt | 2.81 | 8.03 | 0 | 0 | 0.47 | 0 | 0 | 0 | 5.48 | 18.3 | 0.246 |
| | 20 kt | 117.44 | 213.58 | 0 | 0 | 137.80 | 0 | 0 | 55.52 | 0 | 39.7 | 3.430 |
| | 200 kt | 926.74 | 1845.61 | 0 | 0 | 676.29 | 0 | 587.87 | 0 | 0 | 124.0 | 28.408 |

Table E. 5. Unique and no low C case: summary of optimal configuration, $CO_2$ emission rate, NG consumption rate for variable $H_2$ demand.

| | | PV (MW) | B (MWh) | PEM (MW) | Alkaline (MW) | H tank (t $H_2$) | NGCC (MW) | ATR (t $H_2$/day) | SMRCC (t $H_2$/day) | SMR (t $H_2$/day) | $CO_2$ (kt/year) | NG (TBtu/year) |
|---|---|---|---|---|---|---|---|---|---|---|---|---|
| 2030 | 2 kt | 2.67 | 6.21 | 0 | 0 | 14.51 | 0 | 0 | 0 | 6.66 | 18.3 | 0.246 |
| | 20 kt | 29.39 | 68.38 | 0 | 0 | 46.52 | 0 | 0 | 0 | 73.36 | 183.4 | 2.460 |
| | 200 kt | 297.45 | 692.04 | 0 | 0 | 395.38 | 0 | 0 | 0 | 742.41 | 1834.1 | 24.602 |
| 2040 | 2 kt | 2.65 | 6.16 | 0 | 0 | 15.82 | 0 | 0 | 0 | 6.61 | 18.4 | 0.246 |
| | 20 kt | 28.79 | 66.99 | 0 | 0 | 65.38 | 0 | 0 | 0 | 71.87 | 183.4 | 2.460 |
| | 200 kt | 1018.30 | 2369.15 | 0 | 0 | 1220.98 | 0 | 679.64 | 0 | 0 | 124.0 | 28.407 |
| 2050 | 2 kt | 70.37 | 29.90 | 30.25 | 0 | 76.49 | 0 | 0 | 0 | 0 | 0 | 0 |
| | 20 kt | 26.68 | 62.07 | 0 | 0 | 145.09 | 0 | 0 | 0 | 66.59 | 183.5 | 2.461 |
| | 200 kt | 997.74 | 2321.32 | 0 | 0 | 1450.93 | 0 | 665.92 | 0 | 0 | 124.1 | 28.414 |

Table E. 6. Unique and expensive NG case: summary of optimal configuration, $CO_2$ emission rate, NG consumption rate for constant $H_2$ demand.

| | | PV (MW) | B (MWh) | PEM (MW) | Alkaline (MW) | H tank (t $H_2$) | NGCC (MW) | ATR (t $H_2$/day) | SMRCC (t $H_2$/day) | SMR (t $H_2$/day) | $CO_2$ (kt/year) | NG (TBtu/year) |
|---|---|---|---|---|---|---|---|---|---|---|---|---|
| 2030 | 2 kt | 0 | 0 | 0 | 0 | 0 | 0.22 | 0 | 0 | 5.48 | 18.4 | 0.260 |
| | 20 kt | 0 | 0 | 0 | 0 | 0 | 2.19 | 0 | 0 | 54.79 | 184.1 | 2.597 |
| | 200 kt | 0 | 0 | 0 | 0 | 0 | 21.92 | 0 | 0 | 547.95 | 1841.3 | 25.973 |
| 2040 | 2 kt | 0 | 0 | 0 | 0 | 0 | 0.22 | 0 | 0 | 5.48 | 18.4 | 0.260 |
| | 20 kt | 0 | 0 | 0 | 0 | 0 | 2.19 | 0 | 0 | 54.79 | 184.1 | 2.597 |
| | 200 kt | 0 | 0 | 0 | 0 | 0 | 81.96 | 547.95 | 0 | 0 | 151.3 | 33.534 |
| 2050 | 2 kt | 73.12 | 30.69 | 31.26 | 0 | 69.18 | 0.00 | 0 | 0 | 0 | 0 | 0 |
| | 20 kt | 0 | 0 | 0 | 0 | 0 | 2.19 | 0 | 0 | 54.79 | 184.1 | 2.597 |
| | 200 kt | 0 | 0 | 0 | 0 | 0 | 81.96 | 547.95 | 0 | 0 | 151.3 | 33.534 |

Table E. 7. Unique and expensive NG case: summary of optimal configuration, $CO_2$ emission rate, NG consumption rate for variable $H_2$ demand

| | | PV (MW) | B (MWh) | PEM (MW) | Alkaline (MW) | H tank (t $H_2$) | NGCC (MW) | ATR (t $H_2$/day) | SMRCC (t $H_2$/day) | SMR (t $H_2$/day) | $CO_2$ (kt/year) | NG (TBtu/year) |
|---|---|---|---|---|---|---|---|---|---|---|---|---|
| 2030 | 2 kt | 0 | 0 | 0 | 0 | 14.51 | 0.27 | 0 | 0 | 6.66 | 18.4 | 0.260 |
| | 20 kt | 0 | 0 | 0 | 0 | 44.07 | 2.95 | 0 | 0 | 73.64 | 184.1 | 2.597 |
| | 200 kt | 0 | 0 | 0 | 0 | 395.38 | 29.70 | 0 | 0 | 742.4067 | 1.8 | 25.974 |
| 2040 | 2 kt | 75.07 | 31.14 | 31.52 | 0 | 53.61 | 0.00 | 0 | 0 | 0 | 0 | 0 |
| | 20 kt | 0 | 0 | 0 | 0 | 46.52 | 2.93 | 0 | 0 | 73.36 | 184.1 | 2.597 |
| | 200 kt | 0 | 0 | 0 | 0 | 465.24 | 109.73 | 733.57 | 0 | 0 | 151.3 | 33.535 |
| 2050 | 2 kt | 70.37 | 29.90 | 30.25 | 0 | 76.49 | 0.00 | 0 | 0 | 0 | 0 | 0 |
| | 20 kt | 703.66 | 299.01 | 302.54 | 0 | 764.87 | 0.00 | 0 | 0 | 0 | 0 | 0 |
| | 200 kt | 0 | 0 | 0 | 0 | 868.38 | 105.15 | 702.98 | 0 | 0 | 151.3 | 33.536 |

Table E. 8. Base case: summary of optimal configuration, $CO_2$ emission rate, NG consumption rate for constant $H_2$ demand.

| | | PV (MW) | B (MWh) | PEM (MW) | Alkaline (MW) | H tank (t $H_2$) | NGCC (MW) | ATR (t $H_2$/day) | SMRCC (t $H_2$/day) | SMR (t $H_2$/day) | $CO_2$ (kt/year) | NG (TBtu/year) |
|---|---|---|---|---|---|---|---|---|---|---|---|---|
| 2030 | 2 kt | 0 | 0 | 0 | 0 | 0 | 0.22 | 0 | 0 | 5.48 | 18.4 | 0.260 |
| | 20 kt | 0 | 0 | 0 | 0 | 0 | 2.19 | 0 | 0 | 54.79 | 184.1 | 2.597 |
| | 200 kt | 0 | 0 | 0 | 0 | 0 | 81.96 | 547.95 | 0 | 0 | 151.3 | 33.534 |
| 2040 | 2 kt | 0.23 | 0 | 0.05 | 0 | 0.01 | 0.22 | 0 | 0 | 5.47 | 18.4 | 0.258 |
| | 20 kt | 1.59 | 0 | 0 | 0 | 0 | 2.19 | 0 | 0 | 54.79 | 184.0 | 2.576 |
| | 200 kt | 59.39 | 0 | 0 | 0 | 0 | 81.96 | 547.95 | 0 | 0 | 147.1 | 32.747 |
| 2050 | 2 kt | 31.46 | 0 | 12.12 | 6.12 | 2.90 | 2.02 | 0 | 0 | 2.35 | 6.7 | 0.182 |
| | 20 kt | 8.02 | 0 | 0 | 0 | 0 | 10.05 | 0 | 54.79 | 0 | 42.4 | 3.944 |
| | 200 kt | 65.10 | 0 | 0 | 0 | 0 | 81.96 | 547.95 | 0 | 0 | 146.7 | 32.678 |

Table E. 9. Base case: summary of optimal configuration, $CO_2$ emission rate, NG consumption rate for variable $H_2$ demand.

| | | PV (MW) | B (MWh) | PEM (MW) | Alkaline (MW) | H tank (t $H_2$) | NGCC (MW) | ATR (t $H_2$/day) | SMRCC (t $H_2$/day) | SMR (t $H_2$/day) | $CO_2$ (kt/year) | NG (TBtu/year) |
|---|---|---|---|---|---|---|---|---|---|---|---|---|
| 2030 | 2 kt | 3.76 | 0 | 2.67 | 0 | 3.33 | 0.32 | 0 | 0 | 6.17 | 17.1 | 0.241 |
| | 20 kt | 2.47 | 0 | 0 | 0 | 40.21 | 1.97 | 0 | 0 | 74.15 | 184.2 | 2.563 |
| | 200 kt | 95.61 | 0 | 0 | 0 | 395.38 | 72.25 | 742.41 | 0 | 0 | 151.4 | 32.172 |
| 2040 | 2 kt | 21.76 | 0 | 13.78 | 0 | 3.98 | 0.86 | 0 | 0 | 3.66 | 10.4 | 0.179 |
| | 20 kt | 20.62 | 0 | 14.33 | 0 | 46.52 | 2.56 | 0 | 0 | 66.19 | 176.5 | 2.498 |
| | 200 kt | 105.62 | 0 | 0 | 0 | 395.38 | 72.90 | 742.41 | 0 | 0 | 150.2 | 32.086 |
| 2050 | 2 kt | 41.80 | 0 | 21.39 | 3.08 | 19.73 | 2.38 | 0 | 0 | 1.03 | 3.5 | 0.151 |
| | 20 kt | 38.73 | 0 | 21.36 | 0 | 46.52 | 9.81 | 0 | 62.68 | 0 | 41.4 | 3.668 |
| | 200 kt | 108.08 | 0 | 0 | 0 | 465.24 | 74.92 | 733.57 | 0 | 0 | 149.4 | 32.078 |

Table E. 10. No low C electricity case: summary of optimal configuration, $CO_2$ emission rate, NG consumption rate for constant $H_2$ demand.

| | | PV (MW) | B (MWh) | PEM (MW) | Alkaline (MW) | H tank (t $H_2$) | NGCC (MW) | ATR (t $H_2$/day) | SMRCC (t $H_2$/day) | SMR (t $H_2$/day) | $CO_2$ (kt/year) | NG (TBtu/year) |
|---|---|---|---|---|---|---|---|---|---|---|---|---|
| 2030 | 2 kt | 1.31 | 3.65 | 0 | 0 | 0 | 0 | 0 | 0 | 5.48 | 18.4 | 0.246 |
| | 20 kt | 13.07 | 36.47 | 0 | 0 | 0 | 0 | 0 | 0 | 54.79 | 184.1 | 2.460 |
| | 200 kt | 488.69 | 1363.95 | 0 | 0 | 0 | 0 | 547.95 | 0 | 0 | 152.0 | 28.400 |
| 2040 | 2 kt | 1.44 | 3.88 | 0.030 | 0 | 0.01 | 0 | 0 | 0 | 5.48 | 18.4 | 0.246 |
| | 20 kt | 14.30 | 38.75 | 0.247 | 0 | 0.05 | 0 | 0 | 0 | 54.79 | 183.7 | 2.457 |
| | 200 kt | 522.97 | 1421.72 | 0.000 | 0 | 0 | 0 | 547.95 | 0 | 0 | 146.0 | 28.400 |
| 2050 | 2 kt | 34.76 | 14.36 | 15.479 | 0 | 3.00 | 0 | 0 | 0 | 3.05 | 8.3 | 0.106 |
| | 20 kt | 69.61 | 180.43 | 1.499 | 0 | 0.32 | 0 | 0 | 54.75 | 0 | 41.4 | 3.396 |
| | 200 kt | 560.71 | 1459.74 | 0 | 0 | 0 | 0 | 547.95 | 0 | 0 | 141.4 | 28.400 |

Table E. 11. No low C electricity case: summary of optimal configuration, $CO_2$ emission rate, NG consumption rate for variable $H_2$ demand.

| | | PV (MW) | B (MWh) | PEM (MW) | Alkaline (MW) | H tank (t $H_2$) | NGCC (MW) | ATR (t $H_2$/day) | SMRCC (t $H_2$/day) | SMR (t $H_2$/day) | $CO_2$ (kt/year) | NG (TBtu/year) |
|---|---|---|---|---|---|---|---|---|---|---|---|---|
| 2030 | 2 kt | 4.32 | 4.78 | 2.16 | 0 | 4.65 | 0 | 0 | 0 | 6.26 | 17.2 | 0.229 |
| | 20 kt | 14.37 | 33.14 | 1.85 | 0 | 41.05 | 0 | 0 | 0 | 73.12 | 183.1 | 2.447 |
| | 200 kt | 132.05 | 317.64 | 1.72 | 0 | 391.38 | 0 | 0 | 0 | 742.07 | 1840.4 | 24.592 |
| 2040 | 2 kt | 19.93 | 11.83 | 10.98 | 0 | 3.63 | 0 | 0 | 0 | 4.20 | 11.9 | 0.156 |
| | 20 kt | 31.81 | 45.51 | 14.52 | 0 | 46.52 | 0 | 0 | 0 | 66.10 | 176.2 | 2.355 |
| | 200 kt | 537.35 | 1276.26 | 7.30 | 0 | 425.14 | 0 | 734.84 | 0 | 0 | 141.1 | 28.317 |
| 2050 | 2 kt | 44.12 | 18.46 | 20.53 | 0 | 19.94 | 0 | 0 | 0 | 1.99 | 5.7 | 0.073 |
| | 20 kt | 92.63 | 184.13 | 27.96 | 0 | 32.84 | 0 | 0 | 61.34 | 0 | 38.2 | 3.116 |
| | 200 kt | 572.61 | 1326.35 | 31.10 | 0 | 465.24 | 0 | 718.02 | 0 | 0 | 137.0 | 28.059 |

Table E. 12. Expensive NG case: summary of optimal configuration, $CO_2$ emission rate, NG consumption rate for constant $H_2$ demand.

| | | PV (MW) | B (MWh) | PEM (MW) | Alkaline (MW) | H tank (t $H_2$) | NGCC (MW) | ATR (t $H_2$/day) | SMRCC (t $H_2$/day) | SMR (t $H_2$/day) | $CO_2$ (kt/year) | NG (TBtu/year) |
|---|---|---|---|---|---|---|---|---|---|---|---|---|
| 2030 | 2 kt | 0.23 | 0 | 0 | 0 | 0.00 | 0.22 | 0 | 0 | 5.48 | 18.4 | 0.257 |
| | 20 kt | 2.29 | 0 | 0 | 0 | 0.00 | 2.19 | 0 | 0 | 54.79 | 184.0 | 2.570 |
| | 200 kt | 22.94 | 0 | 0 | 0 | 0.00 | 21.92 | 0 | 0 | 547.95 | 1839.8 | 25.698 |
| 2040 | 2 kt | 31.85 | 9.58 | 15.235 | 0 | 2.86 | 0.31 | 0 | 0 | 3.09 | 8.6 | 0.123 |
| | 20 kt | 2.97 | 2.51 | 0 | 0 | 0.00 | 2.06 | 0 | 0 | 54.79 | 184.0 | 2.562 |
| | 200 kt | 111.23 | 93.93 | 0 | 0 | 0.00 | 76.85 | 547.95 | 0 | 0 | 144.6 | 32.225 |
| 2050 | 2 kt | 56.20 | 17.46 | 25.782 | 0 | 5.50 | 0.33 | 0 | 0 | 1.02 | 3.5 | 0.046 |
| | 20 kt | 306.47 | 129.68 | 148.128 | 0 | 23.68 | 0.85 | 0 | 0 | 35.22 | 89.5 | 1.210 |
| | 200 kt | 188.09 | 394.58 | 0 | 0 | 0.00 | 61.92 | 547.95 | 0 | 0 | 141.7 | 31.445 |

Table E. 13. Expensive NG case: summary of optimal configuration, $CO_2$ emission rate, NG consumption rate for variable $H_2$ demand.

| | | PV (MW) | B (MWh) | PEM (MW) | Alkaline (MW) | H tank (t $H_2$) | NGCC (MW) | ATR (t $H_2$/day) | SMRCC (t $H_2$/day) | SMR (t $H_2$/day) | $CO_2$ (kt/year) | NG (TBtu/year) |
|---|---|---|---|---|---|---|---|---|---|---|---|---|
| 2030 | 2 kt | 7.81 | 3.92 | 4.30 | 0 | 2.16 | 0.16 | 0 | 0 | 5.52 | 16.0 | 0.216 |
| | 20 kt | 6.47 | 1.59 | 2.18 | 0 | 38.93 | 1.78 | 0 | 0 | 73.23 | 183.2 | 2.533 |
| | 200 kt | 40.50 | 2.40 | 3.25 | 0 | 387.85 | 17.85 | 0 | 0 | 741.78 | 1840.6 | 25.481 |
| 2040 | 2 kt | 38.60 | 12.06 | 19.44 | 0 | 6.00 | 0.36 | 0 | 0 | 2.37 | 6.8 | 0.099 |
| | 20 kt | 62.04 | 52.78 | 30.59 | 0 | 25.00 | 0.31 | 0 | 0 | 61.09 | 166.0 | 2.221 |
| | 200 kt | 237.30 | 208.27 | 44.19 | 0 | 339.81 | 57.29 | 727.65 | 0 | 0 | 146.7 | 30.853 |
| 2050 | 2 kt | 64.25 | 20.01 | 30.05 | 0 | 16.38 | 0.38 | 0 | 0 | 0.05 | 1.1 | 0.016 |
| | 20 kt | 348.36 | 152.47 | 174.31 | 0 | 99.09 | 0.71 | 0 | 0 | 29.45 | 78.5 | 1.061 |
| | 200 kt | 625.62 | 1181.71 | 181.77 | 0 | 294.25 | 14.73 | 664.88 | 0 | 0 | 132.2 | 27.459 |

Table E. 14. $CO_2$ tax sensitivity analysis case: summary of optimal configuration, $CO_2$ emission rate, NG consumption rate for constant $H_2$ demand.

| | | PV (MW) | B (MWh) | PEM (MW) | Alkaline (MW) | H tank (t $H_2$) | NGCC (MW) | ATR (t $H_2$/day) | SMRCC (t $H_2$/day) | SMR (t $H_2$/day) | $CO_2$ (kt/year) | NG (TBtu/year) |
|---|---|---|---|---|---|---|---|---|---|---|---|---|
| 2 kt | -50 | 27.20 | 0 | 15.38 | 0 | 2.51 | 0.92 | 0 | 0 | 3.16 | 8.9 | 0.159 |
| | base | 31.46 | 0 | 12.12 | 6.12 | 2.90 | 2.02 | 0 | 0 | 2.35 | 6.7 | 0.182 |
| | +50 | 38.46 | 0 | 16.90 | 7.72 | 4.19 | 5.89 | 0 | 0 | 0.00 | 1.6 | 0.233 |
| 20 kt | -50 | 1.69 | 0 | 0 | 0 | 0 | 2.19 | 0 | 0 | 54.79 | 184.0 | 2.575 |
| | base | 8.02 | 0 | 0 | 0 | 0 | 10.05 | 0 | 54.79 | 0 | 42.4 | 3.944 |
| | +50 | 8.17 | 0 | 0 | 0 | 0 | 10.05 | 0 | 54.79 | 0 | 42.4 | 3.942 |
| 200 kt | -50 | 63.23 | 0 | 0 | 0 | 0 | 81.96 | 547.95 | 0 | 0 | 146.9 | 32.700 |
| | base | 65.10 | 0 | 0 | 0 | 0 | 81.96 | 547.95 | 0 | 0 | 146.7 | 32.678 |
| | +50 | 66.65 | 0 | 0 | 0 | 0 | 81.96 | 547.95 | 0 | 0 | 146.6 | 32.661 |

Table E. 15. $CO_2$ tax sensitivity analysis case: summary of optimal configuration, $CO_2$ emission rate, NG consumption rate for variable $H_2$ demand.

| | | PV (MW) | B (MWh) | PEM (MW) | Alkaline (MW) | H tank (t $H_2$) | NGCC (MW) | ATR (t $H_2$/day) | SMRCC (t $H_2$/day) | SMR (t $H_2$/day) | $CO_2$ (kt/year) | NG (TBtu/year) |
|---|---|---|---|---|---|---|---|---|---|---|---|---|
| 2 kt | -50 | 33.49 | 0 | 19.68 | 0 | 19.43 | 1.13 | 0 | 0 | 2.35 | 6.8 | 0.143 |
| | base | 41.80 | 0 | 21.39 | 3.08 | 19.73 | 2.38 | 0 | 0 | 1.03 | 3.5 | 0.151 |
| | +50 | 46.83 | 0 | 24.74 | 3.14 | 18.77 | 3.62 | 0 | 0 | 0 | 1.2 | 0.158 |
| 20 kt | -50 | 28.00 | 1.14 | 18.19 | 0 | 44.07 | 2.64 | 0 | 0 | 64.55 | 174.1 | 2.464 |
| | base | 38.73 | 0 | 21.36 | 0 | 46.52 | 9.81 | 0 | 62.68 | 0 | 41.4 | 3.668 |
| | +50 | 46.83 | 0.16 | 26.79 | 0 | 32.53 | 10.35 | 0 | 61.77 | 0 | 40.6 | 3.616 |
| 200 kt | -50 | 109.77 | 0 | 0 | 0 | 465.24 | 72.13 | 733.57 | 0 | 0 | 150.5 | 32.029 |
| | base | 108.08 | 0 | 0 | 0 | 465.24 | 74.92 | 733.57 | 0 | 0 | 149.4 | 32.078 |
| | +50 | 108.07 | 0 | 0 | 0 | 465.24 | 76.66 | 733.57 | 0 | 0 | 148.6 | 32.100 |

.

**F. Reference.**